\documentclass[aps,prb,preprint,showkeys,superscriptaddress]{revtex4}
\usepackage{amsmath,bm}
\usepackage{graphicx,tabularx}
\usepackage{color}

\begin{document}

\title{Low-Energy Spin Excitation in Coexistent Phase of Antiferromagnetism and $d$-Wave Superconductivity}
\author{Hyun-Jung Lee}
\affiliation{Korea Institute for Advanced Study, 85 Hoegiro, 130-722 Seoul, South Korea}
\author{Tetsuya Takimoto}
\affiliation{Asia Pacific Center for Theoretical Physics, POSTECH, 790-784 Pohang, South Korea}
\date{\today}

\begin{abstract}
{\ Nuclear quadrupole resonance measurements have shown evidences that the heavy fermion compound CeRhIn$_5$ exhibits a coexistent phase 
with commensurate antiferromagnetism and $d$-wave superconductivity. 
In order to clarify the nature of the spin-excitations in the coexistent phase, we have applied the RPA method to an itinerant model, where the effective interaction is given by two mean-field terms of commensurate 
antiferromagnetism and $d$-wave superconductivity. It is shown that, around the transition line between the antiferromagnetic and 
the coexistent states, a low-energy incommensurate spin-excitation is found to develop due to Fermi surface nesting. This feature reminds of the switching of magnetic ordering wave vector observed in the neutron diffraction.  Further, we also calculate spin relaxation rate, which gives a reasonable explanation of the temperature dependence of NQR relaxation rate in the system 
with the coexistent ground state.}
\end{abstract}

\definecolor{darkblue}{rgb}{0,0,0}

\maketitle
\section{Introduction}
\label{sec:intro} 

The strongly correlated electron system shows a rich phase diagram around an antiferromagnetic quantum critical point. The phase diagram consists of a paramagnetic phase 
showing 
a non-Fermi liquid behavior, an antiferromagnetic phase, a d-wave superconducting phase, and a coexistent phase of antiferromagnetism and superconductivity. Actual compounds showing the coexistent phase 
are UPd$_2$Al$_3$\cite{Geibel1991}, CeIn$_3$\cite{Mathur1998}, 
CePd$_2$Si$_2$\cite{Grosche1996}, CeRh$_2$Si$_2$\cite{Movshovich1996,Araki2002}, CeCu$_2$Si$_2$\cite{Steglich1979,Kawasaki2004}, CeRhIn$_5$\cite{Hegger2000,Muramatsu2001}, and { BaFe$_2$(As$_{1-x}$P$_x$)$_2$~\cite{Hashimoto2012}.} Among these compounds, UPd$_2$Al$_3$ with more than two f-electrons is believed that two localized f-electrons contribute to the antiferromagnetism, while remaining f-electrons with a large mass form Cooper pair in the coexistent phase. On the other hand, for other compounds, the same f-electron plays both roles of antiferromagnetism and superconductivity in correlated metal. 

Since the discovery of unconventional superconductivity of CeRhIn$_5$ in 2000, this compound has been investigated intensively by many 
measurements 
under pressure. In the ambient pressure, the compound shows the antiferromagnetic phase around 3.8 K, in which the staggered 
moments 
of 0.6 $\mu_B$/Ce parallel in the a-b plane 
align 
with an ordering wave vector ($\pi, \pi, 0.6\pi$)\cite{Bao2000,Llobet2004,Majumdar2002,Raymond2008}. 
The antiferromagnetic transition temperature decreases with applying pressure. Above 2 GPa, unconventional superconductivity has been observed below 2.1 K.

 In the superconducting state of other compound with the same crystal structure CeCoIn$_5$, the so-called resonance peak has been observed by the inelastic neutron scattering experiment\cite{Stock2008}. A coexistent phase of antiferromagnetism and unconventional superconductivity is also suggested by the nuclear-quadrupole-resonance (NQR) measurement\cite{Yashima2007}. 

Recently, the magnetic neutron diffraction\cite{Aso2009} and NQR measurements\cite{Yashima2009} have been carried out under the pressure to study the compound around the coexistent phase. The magnetic neutron diffraction measurement has showed a change of magnetic ordering wave vector from the original one at the ambient pressure to ($\pi, \pi, 0.8\pi$) in the low temperature region above 1.5 GPa. This implies that an additional spin mode different from the original one at the ambient pressure exists in low energy spectra. Actually, the evidence of this from different measurement is an upturn of NQR relaxation rate devided by temperature $1/(T_1T)$ in the low temperature limit at 1.3 GPa, where the compound will be in the original magnetic state. 
It should be noted that the NQR spectrum in the coexistent phase is reproduced well by an assumption of the commensurate magnetic ordering. 
Further, the value of $1/(T_1T)$ in the low temperature limit in the coexistent phase at 1.5 GPa is two orders larger than that in the superconducting phase at 2.3 GPa, where the quantity becomes temperature independent in the low temperature limit of both phases. 

The main motivation of our study is to clarify the low energy spin excitation around the coexistent phase of antiferromagnetism and unconventional superconductivity. In the antiferromagnetic phase, the antiferromagnetic spin wave appears as a Goldstone mode in the rotationally symmetric spin system. On the other hand, the resonance mode is observed around an antiferromagnetic ordering wave vector in unconventional superconductors. In the coexistent phase, both collective modes are expected. 
Further, recalling the analysis of NQR spectrum in the coexistent phase, 
the reason of the stabilization of the coexistent phase with the commensurate 
magnetic ordering is required. 

This paper is organized as follows. In section~\ref{sec:self-consistent equations}, we introduce the mean-field Hamiltonian with both of the AF exchange and the BCS interaction terms and present the self-consistent solution in the coexisting state of the AF and the SC order. In section~\ref{sec:dynamical spin susceptibility}, we carry out the calculations of the dynamical spin susceptibility in the presence of the AF and the SC long-range order. Section~\ref{sec:local spin dynamics} shows the NMR relaxation rate $1/T_1$ in the coexistence(CO) phase. Section~\ref{sec:conclusion} is a conclusion. We put some technical details in appendixes.

\section{Self-consitent mean-field equations in the coexistence of antiferromagnetism and superconductivity}
\label{sec:self-consistent equations}

In this section, we introduce an effective Hamiltonian describing 
the CO phase of antiferromagnetism and $d$-wave superconductivity 
in a quasi-two-dimentional system, which will be a simplified version of 
CeRhIn$_5$, and show the one electronic state in the CO phase. 
Recalling that the staggered moment is observed in the antiferromagnetic 
state of the compound, the antiferromagnetic phase is a metallic state 
with the Fermi surface. Further, it has been shown by a detail analysis 
of the inelastic neutron scattering data that other compound 
CeCoIn$_5$ with the same crystal structure shows $d_{x^2-y^2}$-
superconductivity at 2.3 K in the ambient pressure\cite{Petrovic2002}. 
Since the superconducting phase of CeRhIn$_5$ appears around 
the antiferromagnetic phase and the transition temperature is quite close 
to that of CeCoIn$_5$, the $d_{x^2-y^2}$-symmetry of superconductivity 
in CeRhIn$_5$ is plausible. 
Assuming the single orbital for electronic state in the unit cell 
of the paramagnetic phase for simplicity, 
the effective Hamiltonian of the system is given by
\begin{equation}
H=H_k+H_{sc}+H_{mg}
\end{equation}
where 
$H_k$, $H_{sc}$, and $H_{mg}$ are the kinetic energy, the interaction of 
$d$-wave pairing, and the antiferromagnetic Heisenberg-type spin exchange terms, 
respectively. 

Applying the tight-binding approximation, 
the non-interacting kinetic term $H_k$ is given as 
\begin{equation}
H_k=\sum_{\bf k}\sum_{\sigma} \varepsilon_{\bf k}c^\dagger_{{\bf k}\sigma}c_{{\bf k}\sigma},
\end{equation}
with the dispersion relation measured from the chemical potential $\mu$
\begin{equation}
\varepsilon_{\bf k}=-2t(\cos k_x + \cos k_y) -4t^\prime \cos k_x \cos k_y-\mu,
\end{equation}
where $c_{{\bf k},\sigma}$ is an annihilation operator of a quasi-particle 
with momentum ${\bf k}$ and spin $\sigma$. 


 In order to describe superconductivity, 
the BCS Hamiltonian is used within the mean-field theory
\begin{eqnarray}
H_{sc}&=&-\frac{1}{2}\sum_{\bf k,k^\prime}\sum_{\sigma}\{ V_{\bf k,k^\prime}\langle c^\dagger_{{\bf k}\sigma}c^\dagger_{{\bf -k}-\sigma}\rangle c_{-{\bf k^\prime}-\sigma}c_{{\bf k^\prime}\sigma} \\
&+& V_{\bf k,k^\prime}\langle  c_{-{\bf k^\prime}-\sigma}c_{{\bf k^\prime}\sigma}\rangle  c^\dagger_{{\bf k}\sigma}c^\dagger_{-{\bf k}-\sigma}\},
\end{eqnarray}
where the 
pairing interaction $V_{\bf k,k^\prime}$ is assumed to take 
following form 
\begin{equation}
V_{\bf k,k^\prime}=V \phi_{\bf k}\phi_{\bf k^\prime}
\label{pairinteraction}
\end{equation}
with 
a basis function of 
$d_{x^2-y^2}$ wave 
gap function 
$\phi_{\bf k}=\cos k_x -\cos k_y$. 
In the following, $\langle\cdots\rangle$ is the thermal average 
with the effective Hamiltonian. 

For the antiferromagnetism, 
the mean-field approximation 
is applied 
for the Heisenberg spin exchange,
\begin{equation}
H_{mg}=-\frac{U}{N_0}\left[ \langle S^z_{-\bf Q} \rangle S^z_{\bf Q} + \langle S^z_{\bf Q} \rangle   S^z_{-\bf Q}\right],
\end{equation}
which is characterized by the staggered magnetic moment,
\begin{equation}
S^{\alpha}_{\bf Q}=
\frac{1}{2}\sum_{\bf k}\sum_{\sigma\sigma^\prime} c^\dagger_{{\bf k}\sigma}\sigma^{\alpha}_{\sigma \sigma^\prime}c_{{\bf k+Q}\sigma^\prime}.
\end{equation}

The order parameters for the superconductivity 
\begin{eqnarray}
{\Delta}_{\bf k}&=&-\frac{1}{2}\sum_{{\bf k^\prime}\sigma}V_{\bf k,k^\prime}{\sigma}\langle c^\dagger_{-{\bf k}-\sigma} c^\dagger_{{\bf k}\sigma}\rangle=\Delta \phi_{\bf k}\label{orderparameter0-1}
\end{eqnarray}
 and the antiferromagnetism
\begin{eqnarray}
m_s&=&\langle S^z_{\bf Q}\rangle
\end{eqnarray}
are self-consistently determined by solving the equations of motion for Green's functions, 
\begin{eqnarray}
\frac{\partial}{\partial \tau}{G({\bf k}, \tau)}&=& -\delta(\tau)-(\varepsilon_{\bf k}-\mu)G({\bf k},\tau)\label{dyson_eq0-1}\\
&+&Um_sF^{z\bf Q}({\bf k},\tau)+\Delta_{\bf k} F^{s\dagger}({\bf k},\tau),\nonumber\\
\frac{\partial}{\partial \tau}{F^{z\bf Q}({\bf k}, \tau)}&=& -(\varepsilon_{\bf k+Q}-\mu)F^{z\bf Q}({\bf k},\tau) \label{dyson_eq0-2}\\
&+&Um_sG({\bf k},\tau)+\Delta_{\bf k+Q} F^{t{\bf Q}\dagger}({\bf k},\tau),\nonumber\\
\frac{\partial}{\partial \tau}{F^{s\dagger}}({\bf k}, \tau)&=& (\varepsilon_{\bf k}-\mu)F^{s\dagger}({\bf k},\tau) \label{dyson_eq0-3}\\
&+&Um_sF^{t\bf Q\dagger}({-\bf k},\tau)+\Delta^*_{\bf k} G({\bf k},\tau),\nonumber\\
\frac{\partial}{\partial \tau}{F^{t{\bf Q}\dagger}}({\bf k}, \tau)&=& (\varepsilon_{\bf k+Q}-\mu)F^{t{\bf Q}\dagger}({\bf k},\tau)\label{dyson_eq1}\\
&+&Um_sF^{s\dagger}({\bf k},\tau)+\Delta^*_{\bf k+Q} F^{z{\bf Q}}({\bf k},\tau),\nonumber
\end{eqnarray}
where the 
Green's functions are defined as
\begin{eqnarray}
G({\bf k}, \tau)&=&-\frac{1}{2}\sum_\sigma  \langle T_{\tau}\left[ c_{{\bf k}\sigma}(\tau) c^\dagger_{{\bf k}\sigma}(0)\right]\rangle, \nonumber\\
F^{z\bf Q}({\bf k}, \tau)&=&-\frac{1}{2}\sum_\sigma  \sigma\langle T_{\tau}\left[ c_{{\bf k+Q}\sigma}(\tau) c^\dagger_{{\bf k}\sigma}(0)\right]\rangle, \nonumber\\
F^{s}({\bf k}, \tau)&=&-\frac{1}{2}\sum_\sigma  \sigma\langle T_{\tau}\left[ c_{{\bf k}\sigma}(\tau) c_{-{\bf k}-\sigma}(0)\right]\rangle, \nonumber\\
F^{t{\bf Q}\dagger}({\bf k}, \tau)&=&-\frac{1}{2}\sum_\sigma  \langle T_{\tau}\left[ c^\dagger_{-{\bf k-Q}-\sigma}(\tau) c^\dagger_{{\bf k}\sigma}(0)\right]\rangle,\nonumber\\ 
\label{define_corr_ftns}
\end{eqnarray}
with $c_{{\bf k}\sigma}(\tau)=e^{H\tau}c_{{\bf k}\sigma}e^{-H\tau}$. $G({\bf k}, \tau)$ is the normal Green's function, while 
$F^{z\bf Q}({\bf k}, \tau)$ and $F^{s}({\bf k}, \tau)$ are the anomalous Green's functions associated to the antiferromagnetism and the superconductivity, respectively. 
The last Green's function $F^{t{\bf Q}\dagger}({\bf k}, \tau)$ 
proportional to the two different order parameters 
takes a non-zero value only in the CO phase, 
and is the so-called $\pi$-triplet pairing. 

Following the symmetry arguments given in Table~\ref{pairsym}, 
the $\pi$-triplet pairing is clarified as an antisymmetric pairing 
in two sublattices. 
Since the pair of two electrons is a wave function of two electrons, 
it is antisymmetric for the exchange of the two electrons. 
In the paramagnetic state, this antisymmetry is satisfied in 
the spin-singlet pairing with even parity or the spin-triplet pairing 
with odd parity. On the other hand, the sublattice index is additionally 
necessary in the phase with the antiferromagnetic order parameter. 
Therefore, the spin-triplet and sublattice-antisymmetric pairing 
with even parity and the spin-singlet and sublattice-antisymmetric pairing 
with odd parity are also possible in the coexistent phase. 
Since the parity is preserved, the spin-triplet and sublattice-antisymmetric 
pairing with even parity is induced in the present case. 

\begin{table}
\vspace{0.4cm}
\begin{center}
\begin{tabular}{|p{2cm}|p{3cm}|p{3cm}|p{3cm}|p{2cm}|}
\hline 
&spin& parity &sublattice\\
\hline
$F^s$&singlet & even & symmetric\\
$F^t$&triplet & odd & symmetric\\
$F^{z\bf Q}$&singlet & odd & antisymmetric\\
$F^{t\bf Q}$&triplet & even & antisymmetric\\
\hline
\end{tabular}
\caption{
Pairings in the coexistent phase of antiferromagnetism and superconductivity.}
\label{pairsym}
\end{center}
\end{table}

The self-consistent mean-field equations for the electron density $n_e$ , the staggered magnetic moment $m_s$ and the superconducting gap function $\Delta_{\bf k}$ are obtained as follows.~\cite{Murakami1998, Aperis2008, Anderson2009}
\begin{eqnarray}
\frac{n_e}{2}&=&\frac{T}{N_0}\sum_{\bf k}\sum_{\omega_n} G({\bf k}, {\rm i}\omega_n)\label{orderparameter0}\\
&=&\frac{1}{N_0}\sum_{\bf k}^\prime \left[1-\frac{E_{{\bf k}+}}{2\tilde{E}_{{\bf k}+}}\tanh\frac{\tilde{E}_{{\bf k}+}}{2T}-\frac{E_{{\bf k}-}}{2\tilde{E}_{{\bf k}-}}\tanh\frac{\tilde{E}_{{\bf k}-}}{2T}\right],\nonumber \\
{m_s}&=&\frac{T}{N_0}\sum_{\bf k}\sum_{\omega_n} F^{z\bf Q}({\bf k}, {\rm i}\omega_n)
\label{orderparameter1}\\
&=&\frac{1}{N_0}\sum_{\bf k}^\prime \frac{Um_s}{\sqrt{t^2_{1\bf k}+(Um_s)^2}}\frac{E_{{\bf k}+}}{2\tilde{E}_{{\bf k}+}}\tanh\frac{\tilde{E}_{{\bf k}+}}{2T}\nonumber\\
&-&\frac{1}{N_0}\sum_{\bf k}^\prime\frac{Um_s}{\sqrt{t^2_{1\bf k}+(Um_s)^2}}\frac{E_{{\bf k}-}}{2\tilde{E}_{{\bf k}-}}\tanh\frac{\tilde{E}_{{\bf k}-}}{2T},\nonumber \\
{\Delta}_{\bf k}&=&{T}\sum_{\bf k^\prime}V_{\bf k,k^\prime}\sum_{\omega_n} F^{s\dagger}({\bf k}, {\rm i}\omega_n)\label{orderparameter2}\\
&=&\sum_{\bf k^\prime}^\prime V_{\bf k,k^\prime}\left[\frac{\Delta_{{\bf k}^\prime}}{2\tilde{E}_{{\bf k^\prime}+}}\tanh\frac{\tilde{E}_{{\bf k^\prime}+}}{2T}+\frac{\Delta_{{\bf k^\prime}}}{2\tilde{E}_{{\bf k^\prime}-}}\tanh\frac{\tilde{E}_{{\bf k^\prime}-}}{2T}\right], \nonumber
\end{eqnarray}
where $\sum_{\bf k}^\prime$ is the momentum summation within the magnetic Brillouin zone.  
Here, the dispersion relations of quasi-particle in the coexistent phase 
is obtained by solving the secular equation of four Green's functions 
as follows,
\begin{eqnarray}
\tilde{E}^2_{\bf k\pm}&=&\left[t_{2\bf k}-\mu\pm\sqrt{t^2_{1\bf k}+(Um_s)^2}\right]^2+|\Delta_{{\bf k}}|^2\nonumber \\
&=&{E}^2_{\bf k\pm}+|\Delta_{{\bf k}}|^2\label{dispersion_CO}
\end{eqnarray}
where 
\begin{eqnarray}
{E}_{\bf k\pm}=t_{2\bf k}-\mu\pm\sqrt{t^2_{1\bf k}+(Um_s)^2}\label{dispersion_AF}
\end{eqnarray}
are the excitation-energy in the antiferromagnetic state with
${t}_{1\bf k}=-2t(\cos{k_x}+\cos{k_y})$ and
${t}_{2\bf k}=-4t^\prime\cos{k_x}\cos{k_y}$.
The detailed calculations for Eq.~(\ref{orderparameter0})$\sim$(\ref{dispersion_AF}) are presented in Appendix~\ref{app:dispersion}.
{\renewcommand{\figurename}{}\renewcommand{\thefigure}{Fig. 1 (Color online)} 
\begin{figure}
\begin{center}
\vspace{0.4cm}
  \includegraphics[angle=0,width=0.45\textwidth]{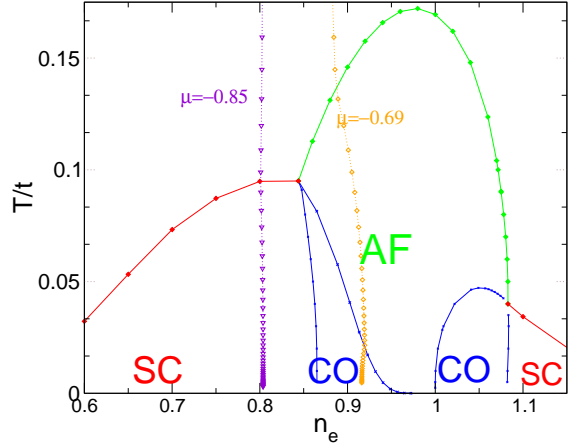}
\end{center}
\vspace{-0.5cm}
\caption{Phase diagram obtained by solving the self-consistent equations, Eq.~(\ref{orderparameter1}) and Eq.~(\ref{orderparameter2}) with a set of parameters $U=2.2$, $V=0.8$ and $t^\prime=-0.2$ in units of the nearest neighber hopping $t$.  The trajectory with open symbol shows the temperature evolution of the system for the fixed chemical potential $\mu=-0.69$ and $\mu=-0.85$.}
\label{pb}
\end{figure}}
{\renewcommand{\figurename}{}\renewcommand{\thefigure}{Fig. 2 (Color online)} 
 \begin{figure}
\begin{center}
\vspace{0.4cm}
  \includegraphics[angle=0,width=0.5\textwidth]{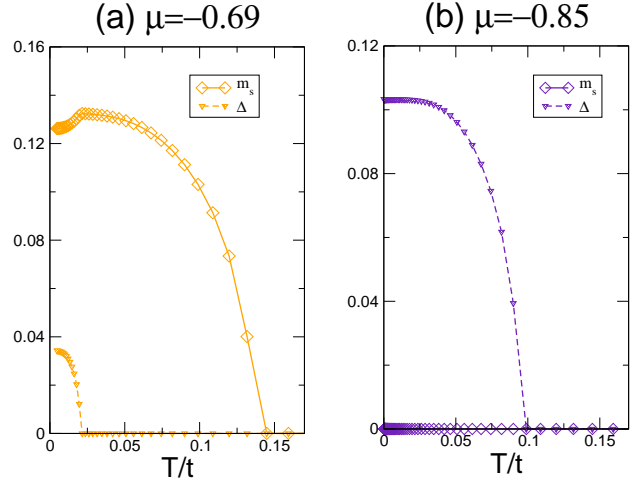}
\end{center}
\vspace{0.cm}
\caption{
The AF and the SC order parameters $m_z$ and $\Delta$ for the chemical potential $\mu=-0.69$ and $\mu=-0.85$ in units of the nearest neighbor hopping $t$. At $\mu=-0.69$, the average magnetization is slightly suppressed when the SC long-range order coexists with the AF.}
\label{op}
\end{figure}
}
The phase diagram of the system with a parameter set $t'/t=-0.2$, 
$U/t=2.2$, and $V/t=0.8$ is shown in Fig. 1 by solving the self-consistent mean-field equations. Each phase boundary in Fig. 1 is obtained with tunning $T$ or $n_e$ around the critical values, around which Eq.~(\ref{orderparameter1}) and Eq.~(\ref{orderparameter2}) can be linearized with respect to the small order parameters $m_s$ and $\Delta_{\bf k}$. For examples, the phase boundary between the CO and the pure SC phase can be tunned by the linearized equation,
\begin{eqnarray}
1&=&\frac{U}{2N_0}\sum_{\bf k}^\prime \frac{1}{|t_{1\bf k}|}\frac{\nu_{\bf k+}}{\sqrt{\nu_{\bf k+}^2+|\Delta_{\bf k}|^2}}\tanh\frac{\sqrt{\nu_{\bf k+}^2+|\Delta_{\bf k}|^2}}{2T} \nonumber\\
&-&\frac{U}{2N_0}\sum_{\bf k}^\prime \frac{1}{|t_{1\bf k}|}\frac{\nu_{\bf k-}}{\sqrt{\nu_{\bf k-}^2+|\Delta_{\bf k}|^2}}\tanh\frac{\sqrt{\nu_{\bf k-}^2+|\Delta_{\bf k}|^2}}{2T} \nonumber\\
\label{CO_instab1}
\end{eqnarray}
where $\nu_{\bf k\pm}=t_{2\bf k}-\mu\pm|t_{1\bf k}|$.  On the other hand, the phase boundary between the CO and the pure AF phase can be tunned by the linearized equation,
\begin{eqnarray}
1&=&\sum_{\bf k}V\phi_{\bf k}^2\left[\frac{1}{2{E}_{{\bf k}+}}\tanh\frac{{E}_{{\bf k}+}}{2T}+\frac{1}{2{E}_{{\bf k}-}}\tanh\frac{{E}_{{\bf k}-}}{2T}\right], \nonumber\\
\label{CO_instab1}
\end{eqnarray}
where 
$E_{\bf k\pm}$ is the excitation energy in the AF phase defined in Eq.~(\ref{dispersion_AF}).


In the coexistent phase, one order parameter is affected by another 
order parameter. 
The effect of $\Delta\neq 0$ on the magnetic order parameter $m_s$ 
in the CO phase is shown in Fig. 2(a) with the fixed chemical potential 
at $\mu/t=-0.69$, whose trajectory is also shown in the phase diagram. 
The existence of $\Delta\neq 0$ reduces $m_s$ below $T<T_c$ 
as shown in Fig. 2(a). On the other hand, a similar figure of 
$\Delta$ in the pure SC state with $\mu/t=-0.85$ of 
the phase diagram shows usual behavior of mean-field order parameter. 
Although the suppression of $m_s$ in the CO phase is less than 10 percent 
of the full magnetism in the pure AF phase, 
the SC gap opening at the Fermi surface makes a significant change 
into the density of states near the Fermi level 
as shown in Fig. 3(b) and Fig. 4(b).
{\renewcommand{\figurename}{}\renewcommand{\thefigure}{Fig. 3 (Color online)} 
   \begin{figure}
      \begin{center}
        \vspace{0.2cm}
        \begin{tabular}{cc}
          \includegraphics[angle=0,width=0.35\textwidth]{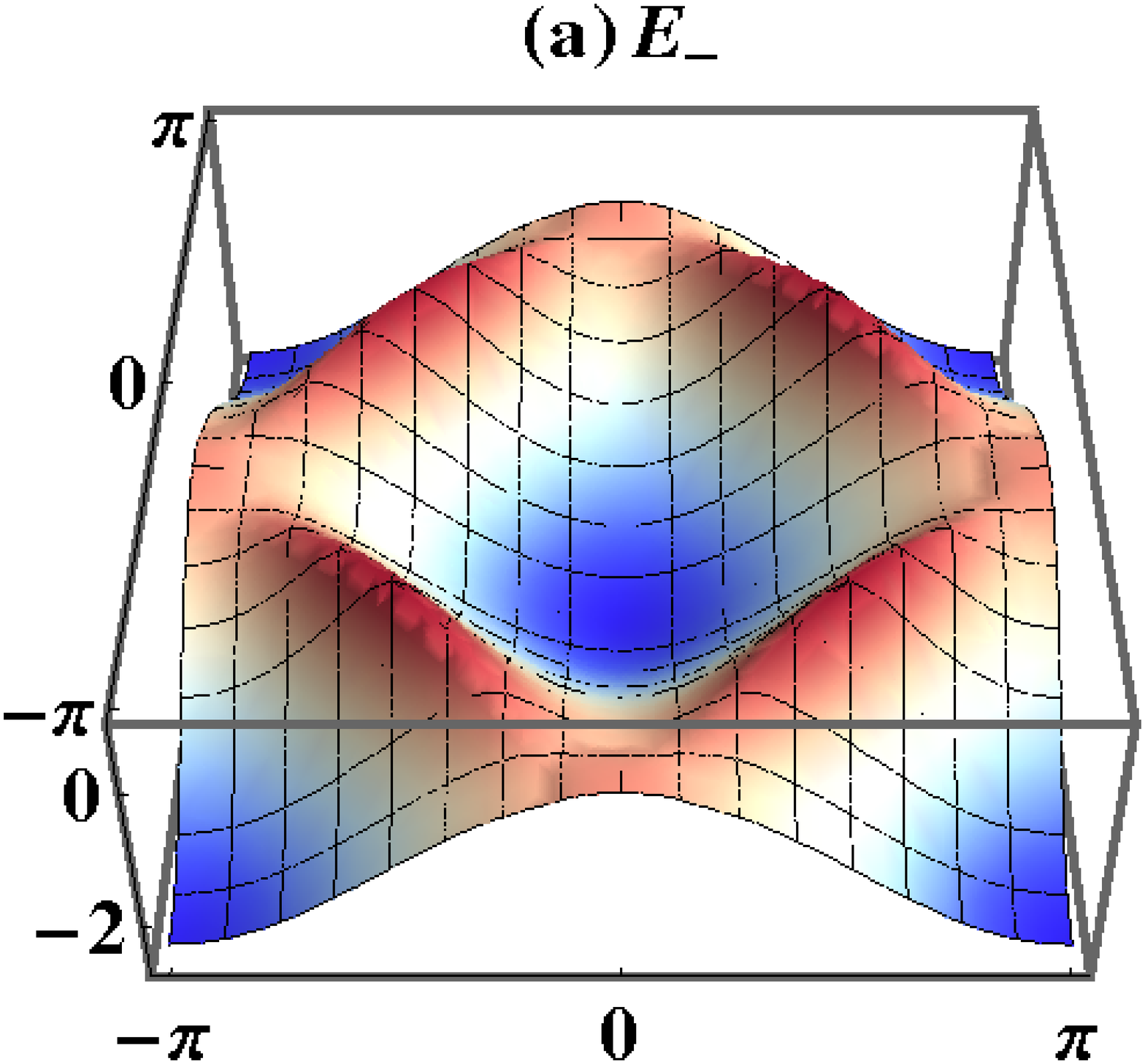}&
          \includegraphics[angle=0,width=0.28\textwidth]{fig3b.eps}\\
          \includegraphics[angle=0,width=0.3\textwidth]{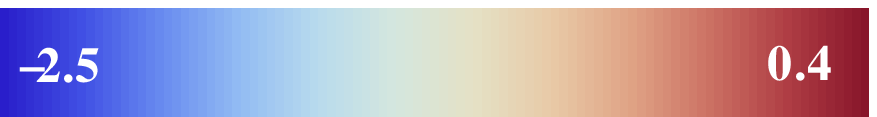}&
        \end{tabular}
      \end{center}
      \vspace{0.0cm}
      \caption{(a) 
A dispersion relation 
${E}_{\bf k-}=t_{2\bf k}-\mu-\sqrt{t^2_{1\bf k}+(Um_s)^2}$ 
crossing the Fermi level 
with $m_s=0.132$, $T=0.0347$ and $\mu=-0.69$ 
in the AF state.  
 (b) Density of states $\rho(E)$ of the $E_{\bf k-}$-band in the AF phase.}
      \label{fs_af}
    \end{figure}
}
{\renewcommand{\figurename}{}\renewcommand{\thefigure}{Fig. 4 (Color online)} 
    \begin{figure}
      \begin{center}
        \vspace{0.2cm}
        \begin{tabular}{cc}
          \includegraphics[angle=0,width=0.35\textwidth]{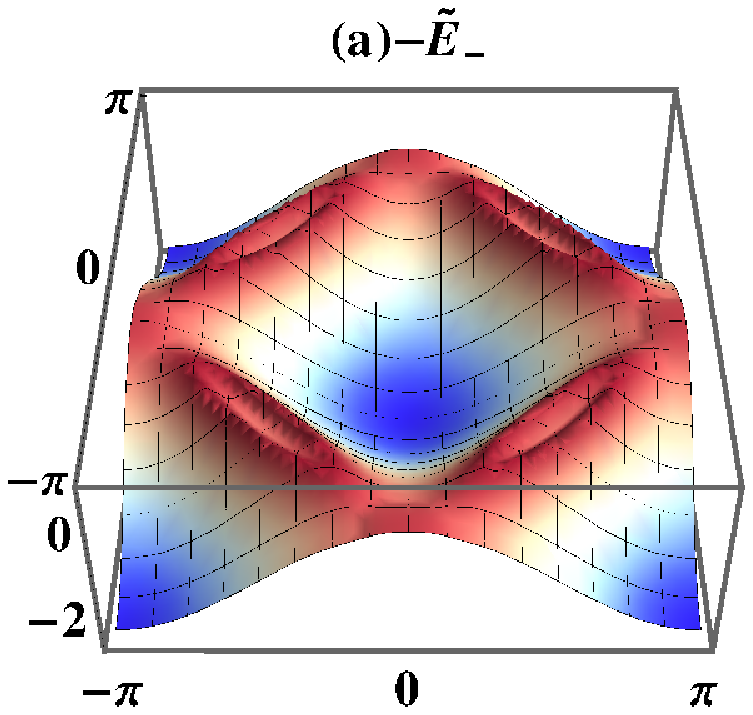}&
          \includegraphics[angle=0,width=0.28\textwidth]{fig4b.eps}\\
          \includegraphics[angle=0,width=0.3\textwidth]{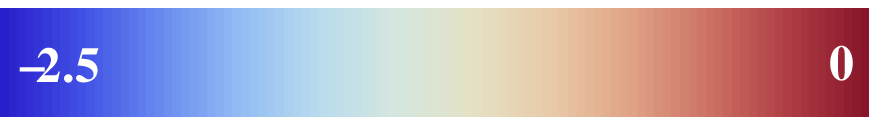}&\\

        \end{tabular}
      \end{center}
      \vspace{0.0cm}
      \caption{(a) 
An energy branch 
$-\tilde{E}_{\bf k-}=-\sqrt{E_{\bf k-}^2+|\Delta_{\bf k}|^2}$ 
below the Fermi level 
in the CO phase with $m_s=0.132$, $\Delta=0.012$, $T=0.0196$ and $\mu=-0.69$. 
(b) Density of states $\rho(E)$ of the $-\tilde{E}_{\bf k-}$-band 
in the CO phase.}
      \label{fs_co}
    \end{figure}
}

Fig. 3(a) and Fig. 4(a) show the dispersion of the excitation energy in the AF and the CO phases in the 2D Brillouin  zone (BZ), respectively. For convenience, we plot only $E_{\bf k-}$ and $-\tilde{E}_{\bf k-}$ bands located near the Fermi level. In the AF phase(Fig. 3(a)), there are four Fermi surfaces with $E_{\bf k -}=0$  around the magnetic BZ boundary, $|k_x|+|k_y|=\pi$, each of which encompasses a hole-pocket with $E_{\bf k -}>0$. In the CO phase(Fig. 4(a)), the Fermi surface is split by the SC gap $\Delta_{\bf k}$, whose maximum $\Delta^{max}_{\bf k}=0.024$ is less than one percent of the band width $W\sim 2.5$. We see that the SC gap opening causes sharp coherence peaks near the Fermi level, with which the spin-excitation spectrum exhibits considerable enhancement by a nesting effect, as shown in Fig. 5(b). We discuss the issue in the following section in terms of the dynamical spin susceptibility calculated with the RPA.



\section{Dynamical spin susceptibility in the presence of antiferromagnetic and superconducting long-range order}
\label{sec:dynamical spin susceptibility}

{\renewcommand{\figurename}{}\renewcommand{\thefigure}{Fig. 5 (Color online)} 
    \begin{figure}
      \begin{center}
        \vspace{0.2cm}
        \noindent
        \begin{tabular}{cc}
          \includegraphics[angle=0,width=0.34\textwidth]{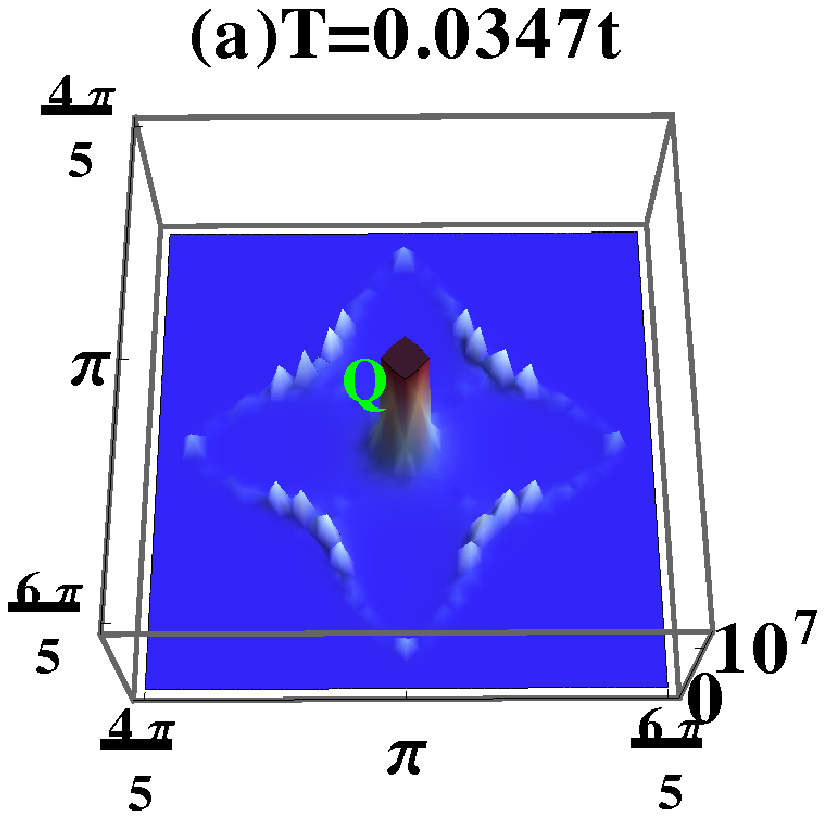}&
          \includegraphics[angle=0,width=0.34\textwidth]{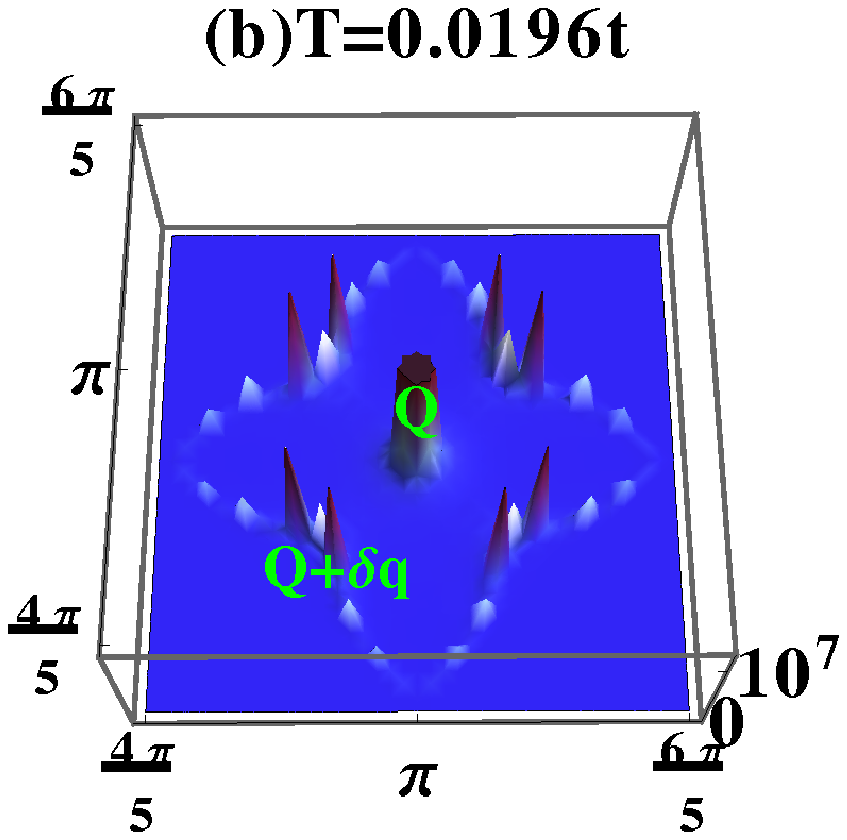}\\
          \includegraphics[angle=0,width=0.34\textwidth]{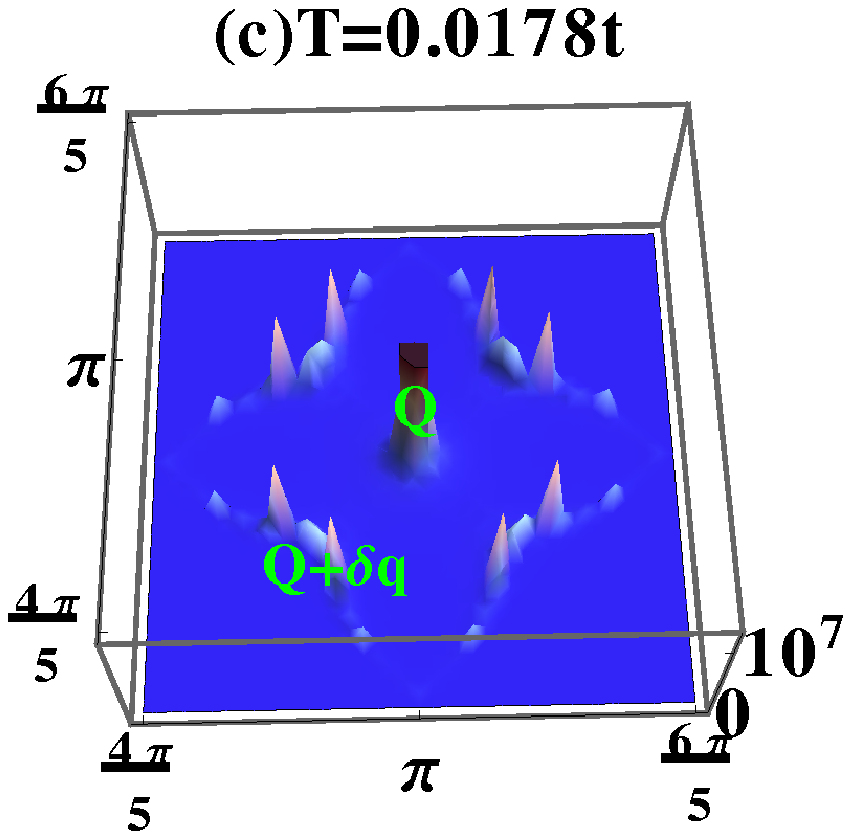}&
          \includegraphics[angle=0,width=0.34\textwidth]{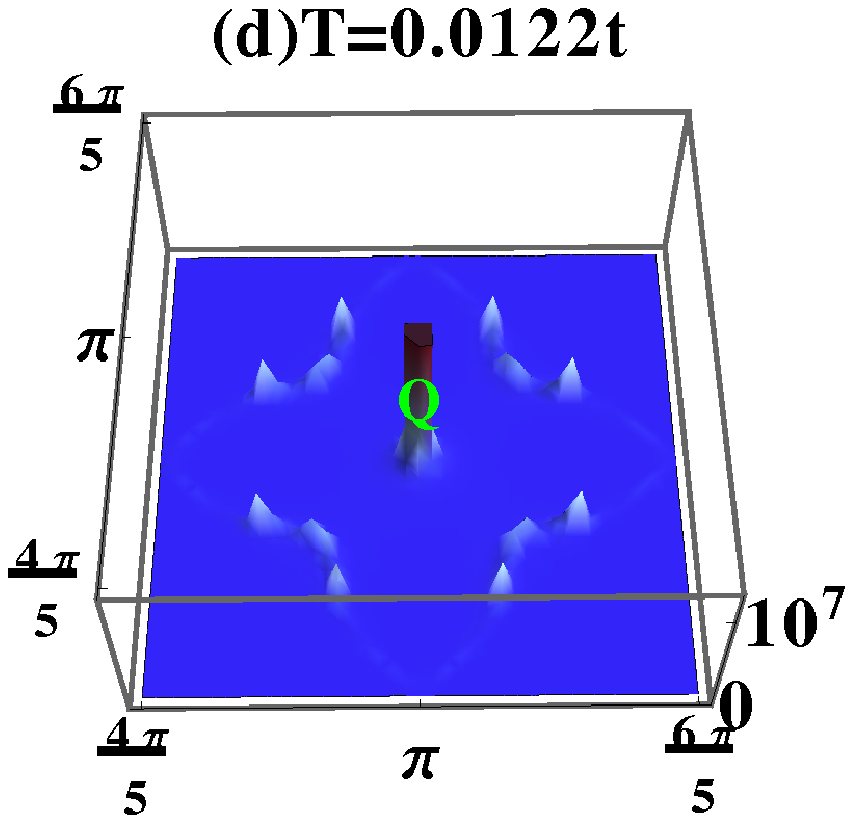}
        \end{tabular}
        \begin{tabular}{c}
          \includegraphics[angle=0,width=0.45\textwidth]{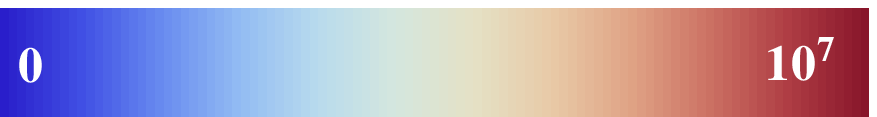}
        \end{tabular}
      \end{center}
      \vspace{0.2cm}
\caption{The momentum-dependence of the spin-excitation spectrum Im$\chi^{\perp}_{RPA}({\bf q}, \omega)/\omega$ in the limit of $\omega\rightarrow 0$ after the analytic continuation ${\rm i}\Omega_n\rightarrow\omega+{\rm i}\delta$ at four different temperatures $T=$(a)$0.0347$, (b)$0.0196$, (c)$0.0178$ and (d)$0.0122$ in units of the nearest neighbor hopping coefficient $t$. The chemical potential is fixed at $\mu=-0.69$, with which superconductivity develops at $T=T_c=0.0196$.  The singularity at {\bf q=Q} appearing at all temperature is the Goldstone mode due to the AF long-range order. (b) At the critical temperature $T=T_c$, the spin-excitation spectrum exhibits another incommensurate peaks at {\bf q=Q+$\delta$q} with $\delta {\bf q}/\pi=(\pm0.08,\pm0.06)$ and $(\pm0.06,\pm0.08)$. The momentum of the incommensurate peaks agrees with the nesting vector, as shown in Fig. 6.}
      \label{ins}
    \end{figure}
}
In order to study the magnetic correlation around the coexistent phase, 
we calculate dynamical spin susceptibility. 
 In the AF and the CO state with $m_s\neq 0$, there are two sites of different sub-lattices in the magnetic unit cell. Then, the area of magnetic Brillouin  zone is a half of the original one, and the ordering wave vector {\bf Q} becomes a reciprocal lattice vector. Therefore, 
momenta of initial and final states of relevant spin fluctuation can differ by the reciprocal lattice vector as the Umklapp process. 
In this case, the spin fluctuations have off-diagonal terms in momentum space representation. 
Since this fact is independent of components of spin operators, 
we introduce a generalized form of the fluctuation with $\alpha, \beta=0, x,y,z,$ components\cite{Ueda1978}, as follows,
\begin{equation}
\chi^{\alpha\beta}({\bf q},{\bf q^\prime},{\rm i} \Omega_n)=\int_0^{1/T} d\tau e^{{\rm i} \Omega_n \tau}\langle T_\tau\left[ \delta S^\alpha_{\bf q}(\tau)\delta S^\beta_{-\bf q^\prime}(0)\right]\rangle
\end{equation}
where $S^0_{\bf q}=\frac{1}{2}\sum_{\bf k}\sum_\sigma c^\dagger_{{\bf k}\sigma}c_{{\bf k+q}\sigma}$ is the charge operator with a momentum {\bf q}. 
 
In a phase with a staggered moment $m_s\neq 0$ parallel to $z$-direction, the transversal components of $\chi^{\alpha\beta}({\bf q,q^\prime,} {\rm i\Omega_n})$ with $\alpha,\beta=x,y$ 
are decomposed from the charge and longitudinal components. 
The transversal components 
have the contribution of spin-waves 
dominating the low-energy spin excitation, while the longitudinal component of $\chi^{zz}({\bf q,q^\prime,} {\rm i\Omega_n})$ reduces to be massive due to the staggered moment.
Especially, when the system is rotationally invariant, as in the present case, the spin wave becomes a Goldstone mode at the ordering wave vector {\bf Q}.  
%
 Therefore the spin-excitation spectrum in the presence of the magnetic long-range order, which can be probed by the inelastic neutron scattering (INS) and the nuclear quadrupole resonance (NQR) experiments, 
is dominated by 
the transverse component $\chi^{\perp}({\bf q}, {\rm i\Omega_n})$, 
especially, around the superconducting instability in the AF state. 
{\renewcommand{\figurename}{}\renewcommand{\thefigure}{Fig. 6 (Color online)} 
\begin{figure}
\begin{center}
\vspace{0.2cm}
  \includegraphics[angle=0,width=0.55\textwidth]{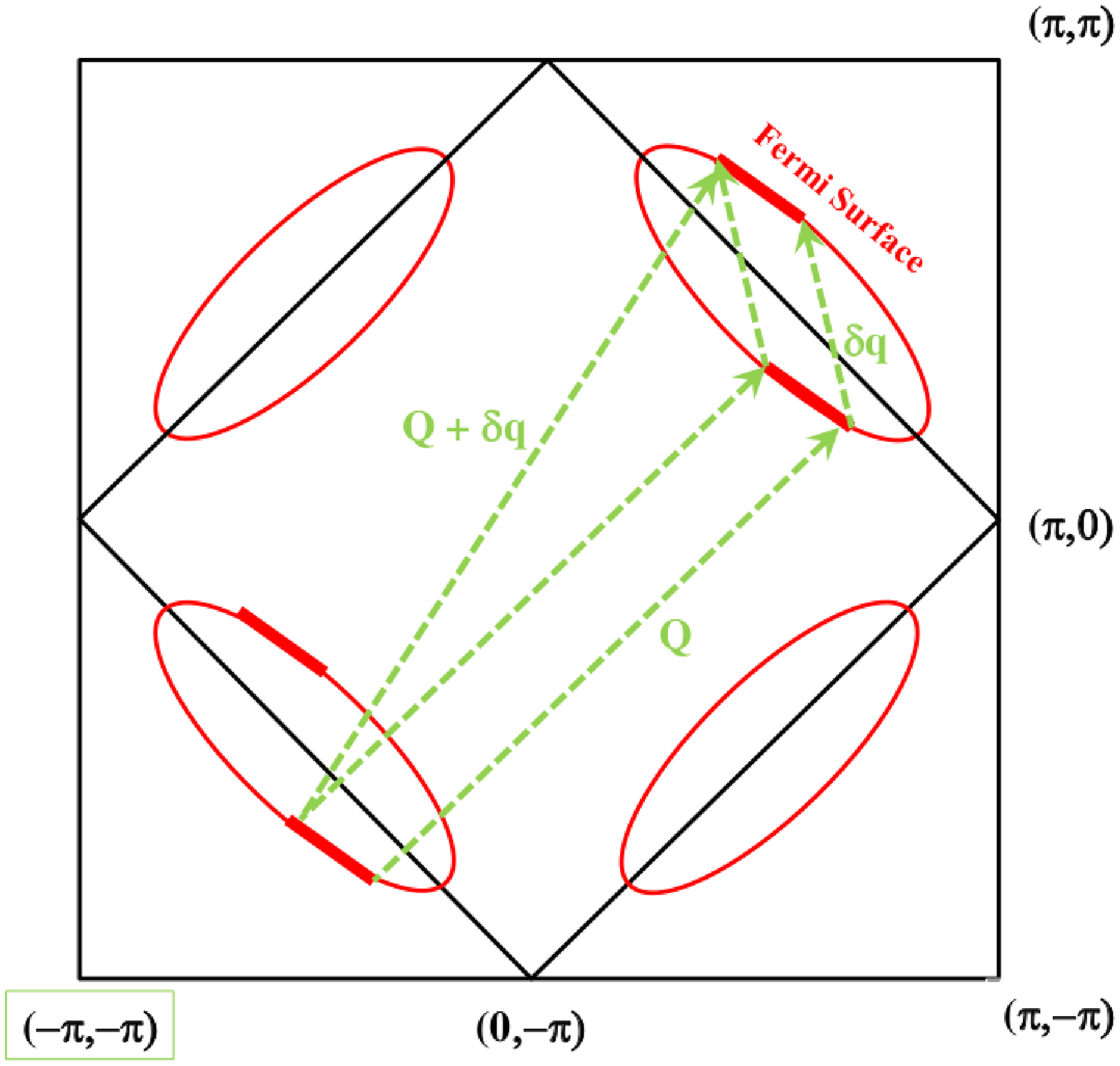}
\end{center}
\vspace{0.0cm}
\caption{Nesting vector $Q+\delta q$}
\label{nesting}
\end{figure}
}
In the RPA calculations of dynamical susceptibility in the CO phase, the transversal component of the RPA susceptibility $\chi^{\perp}_{RPA}({\bf q}, {\rm i\Omega_n})$ is written as a $4\times 4$ matrix consisting of the transverse component of the irreducible susceptibility $\bar{\chi}^{\alpha\beta}({\bf q,q^\prime,} {\rm i\Omega_n})$ with $\alpha,\beta=x,y$ as described in Appendix~\ref{app:spin fluctuation} in detail. 
Even within RPA, the Goldstone mode appears in both AF and CO phases. { The equality of the divergence of transverse susceptibility with the self-consistent equation of the staggered moment in Eq.~(\ref{CO_instab1}) is analytically shown in the coexistent state of our model, as in the antiferromagnetic case~\cite{Ueda1978}. }

Figure 5 shows the momentum-dependence of the spin-excitation spectrum Im$\chi^{\perp}_{RPA}({\bf q}, \omega)/\omega$ in the limit of $\omega\rightarrow 0$ at four different temperatures. The chemical potential is fixed at $\mu=-0.69$, with which a system shows transitions to the AF and the CO states at $T_N=0.132$ and $T_c=0.0196$, respectively. 
In Fig. 5, every temperature is chosen below $T_N$ so that one can observe 
the Goldstone mode at {\bf Q}=$(\pi,\pi)$
\begin{equation}
 \lim_{\omega\rightarrow 0}\lim_{\bf q\rightarrow Q}{\rm Im}\chi^{\perp}_{RPA}({\bf q}, \omega)=\infty.
\end{equation}
%
At $T=T_c=0.0196$, where $d$-wave superconductivity starts to coexist with the AF, we find sharp enhancement of spin excitation spectrum at ${\bf q=Q+\delta q}$ with $\delta {\bf q}/\pi=(\pm0.08,\pm0.06)$ and $(\pm0.06,\pm0.08)$. 
 The incommensurate spin excitation at {\bf q=Q+$\delta$q} turns out to be a nesting contribution as shown in { Fig. 6~\cite{Rowe2012}.}
Such peaks appear only in the vicinity of $T=T_c$ (Fig. 5(b)) and are rapidly suppressed with lowering $T$ below the transition temperature (Fig. 5(c),5(d)), because the electronic states around the Fermi level is gapped in the SC state. {\color{darkblue} Decreasing temperature below $T=T_c$, the frequency of incommensurate spin resonance shifts from around zero to a finite value smaller
than $2\Delta$, as shown in Fig. 7(c) and 7(d). Recalling the spin resonance stabilizes the $d$-wave superconducting phases of high-$T_c$ cuprates and CeCoIn$_5$, the coexistent state of our model will be stabilized by the incommensurate spin resonance, which is characteristic of the coexistent phase. }
{\renewcommand{\figurename}{}\renewcommand{\thefigure}{Fig. 7 (Color online)} 
    \begin{figure}
      \begin{center}
        \vspace{0.2cm}
        \noindent
        \begin{tabular}{cc}
          \includegraphics[angle=0,width=0.3\textwidth]{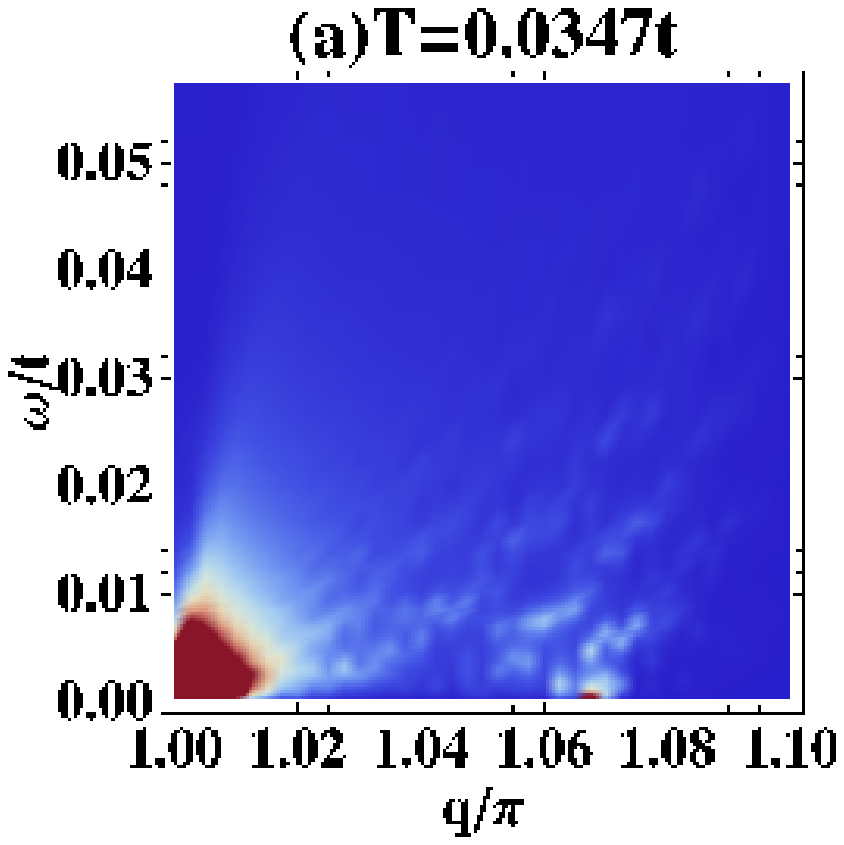}&
          \includegraphics[angle=0,width=0.3\textwidth]{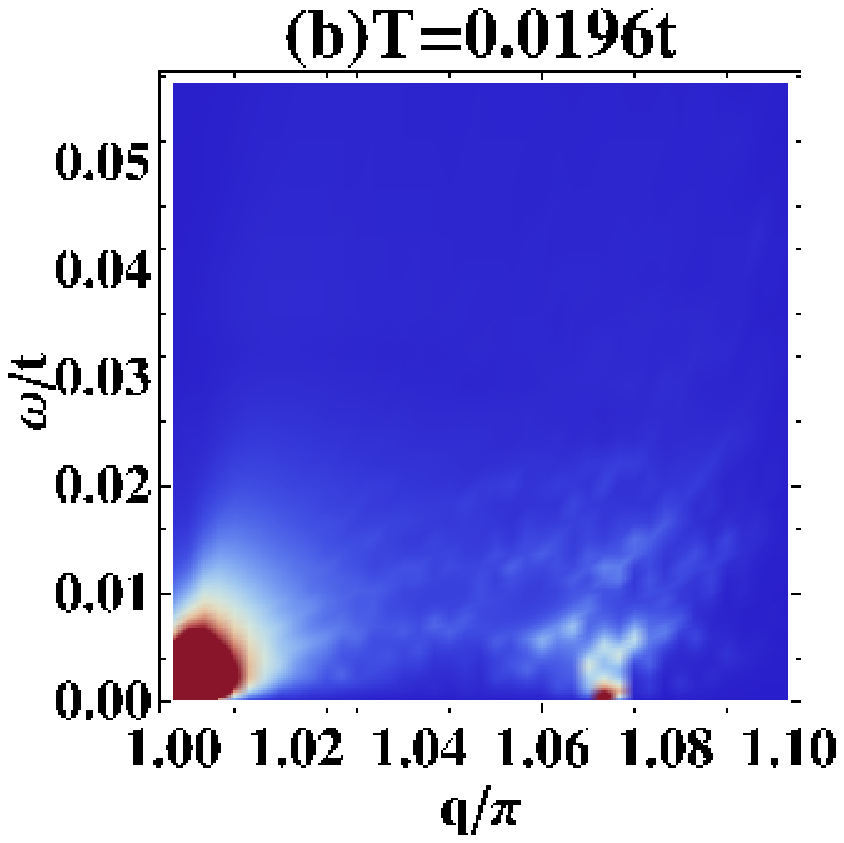}\\
          \includegraphics[angle=0,width=0.3\textwidth]{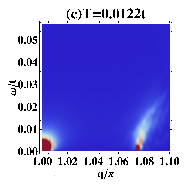}&
          \includegraphics[angle=0,width=0.3\textwidth]{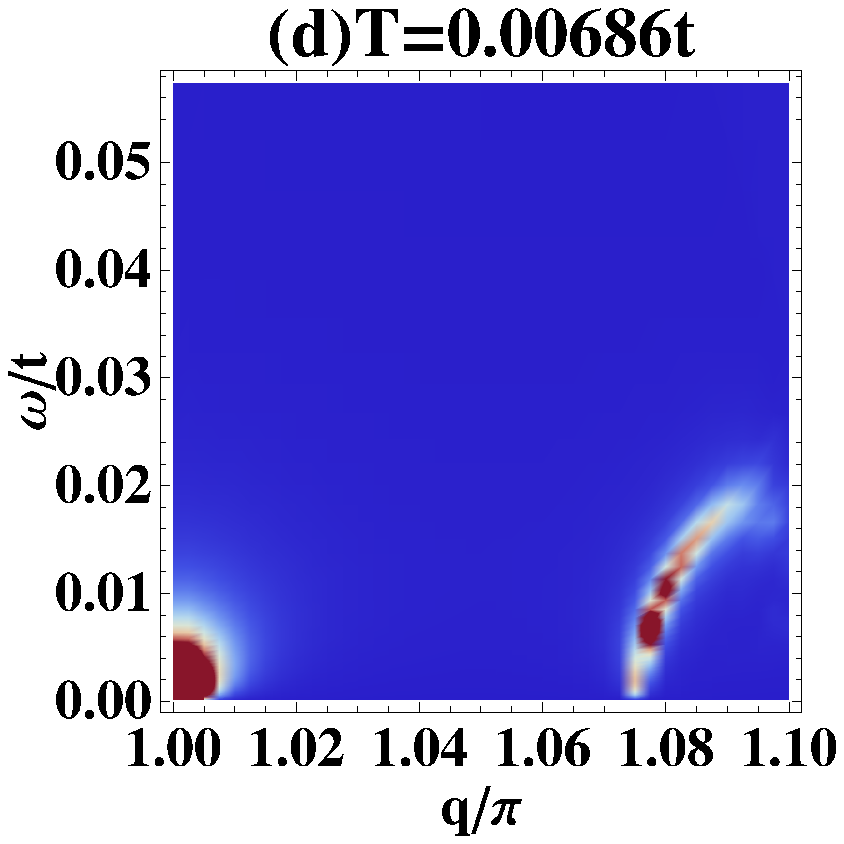}
        \end{tabular}
        \begin{tabular}{c}
          \includegraphics[angle=0,width=0.4\textwidth]{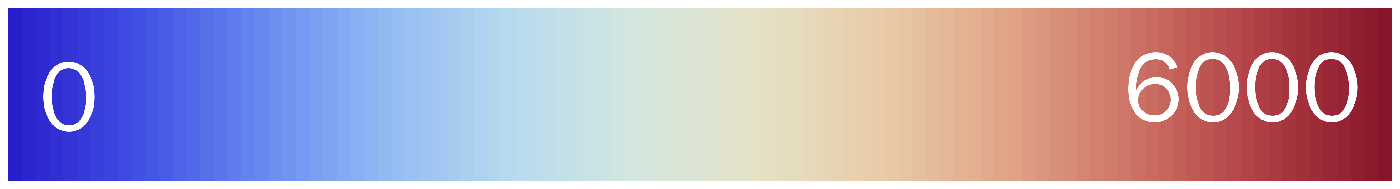}
        \end{tabular}
      \end{center}
      \vspace{0.2cm}
\caption{{\color{darkblue} The spin-excitation spectrum Im$\chi^{\perp}_{RPA}({\bf q}, \omega)$ in the $q$-$\omega$ plane along the diagonal direction $q_x=q_y$ at four different temperatures $T=$(a)$0.0347$, (b)$0.0196$, (c)$0.0122$ and (d)$0.00686$ in units of the nearest neighbor hopping coefficient $t$. The chemical potential is fixed at $\mu=-0.69$, with which superconductivity develops at $T=T_c=0.0196$.  The strong signal at {\bf q=Q} appearing at all temperature is the Goldstone mode due to the AF long-range order. The incommensurate zero-energy mode appearing at $T=T_c$ in Fig. 5(b) and Fig. 7(b) converts to a strong resonance peak as temperature decreases below $T=T_c$ (Fig. 7(c) and 7(d)).} }
      \label{fig_disp}
    \end{figure}
}
In the following, we compare this result with experimental data of CeRhIn$_5$ 
to give a comment. The neutron diffraction experiment shows that the ordering wave vector approaches from an incommensurate one at the ambient pressure $(\pi,\pi,0.6\pi)$ to another incommensurate one $(\pi,\pi,0.8\pi)$ closer to the commensurate one at 1.48 GPa\cite{Aso2009}. Similarly, it has been reported that the NQR spectrum also change a broad shape interpreted by assuming an incommensurate order at the ambient pressure to a sharp peak form explained by the commensurate order at 1.86 GPa, where this compound is in the coexistent phase in the low temperature region\cite{Yashima2009}. On the other hand, it has been shown by the dHvA experiment that the $f$-electronic character also changes from the localized one at ambient pressure to an itinerant one above 2.3GPa due the Kondo effect\cite{Shishido2005}. Considering these, switching of the ordering wave vector from the incommensurate to the commensurate one seems to be attributed to the Kondo coupling between the $f$- and conduction-electrons, which changes the shape of the Fermi surface of the quasi-particles. {\color{darkblue} Now, the sharp incommensurate spin excitation intensity at $T=T_c$
shown in Fig. 5(b) means an instability of the commensurate
antiferromagnetism, though the order parameter is of the commensurate
antiferromagnetism. Around the instability, the free energy of
corresponding incommensurate antiferromagnetic state may be lower than
that of the commensurate one so that an first-order phase transition
from the commensurate to the incommensurate ones is expected in our
model. Then, the incommensurate antiferromagnetic state will have some
similarity with the incommensurate ordered state observed by the neutron
diffraction measurement under the pressure, because heavy
quasi-particles around the Fermi level will be responsible for the
incommensurate antiferromagnetism, where the itinerant picture using
quasi-particles will be applicable. In addition, it is desirable to
observe the spin resonance mode characteristic of the coexistent phaes
as shown in Fig. 7.}

\begin{figure}
\begin{center}
\vspace{0.5cm}
  \includegraphics[angle=0,width=0.5\textwidth]{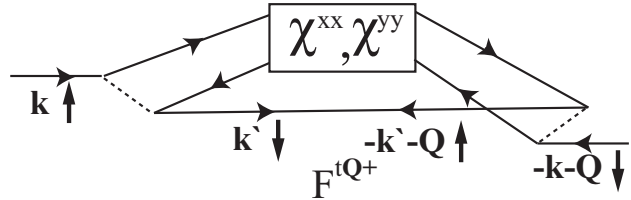}
\end{center}
\vspace{0.0cm}
\caption{Feynman diagram of $\Delta F^{t{\bf Q}\dagger}(k)$. The dashed line means the on-site Coulomb repulsion U.}
\label{FtQ1}
\end{figure}

{\color{darkblue} Finally, we give another consideration on how the coexistent phase with the commensurate magnetic order is stabilized.} As the stability of coexistent phase has been already discussed, the sign of a coupling term, like $\gamma|\Delta_{\bf k}|^2m_s^2+\bar{\gamma}\Delta_{\bf k}\Delta_{\bf k+Q}m_s^2$ between two types of order parameters in the free energy, determines whether the coexistent state is favorable (--sign) or not (+sign)\cite{Sigrist1991}. Similarly, the magnitude of the $\pi$-triplet pairing $F^{t{\bf Q}\dagger}({\bf k}, \tau)$ will give an important information, because the $\pi$-triplet pairing is proportional to the product of two different order parameters within the mean-field theory. Beyond the mean-field theory, a diagram shown in~\ref{FtQ1} contributes to the $\pi$-triplet pairing as follows,
\begin{eqnarray}
 &&\Delta F^{t{\bf Q}\dagger}(k)
  =-G^{(0)}({\bf k+Q},-{\rm i}\omega_n)G({\bf k},{\rm i}\omega_n)\times\nonumber\\
 &&\times U^2\frac{T}{N_0}\sum_{k'}(\chi^{xx}(k-k')+\chi^{yy}(k-k'))
    F^{t{\bf Q}\dagger}(k').
\end{eqnarray}
Since the spin fluctuation exchange term includes the spin-wave mode, 
the correction is expected to enhance the mean-field value of 
$F^{t{\bf Q}\dagger}(k)$. Through the correlation processes as given here, the coexistent state with the commensurate magnetic order will be stabilized. Actual calculation is remained as a future problem. 



\section{Spin-lattice relaxation rate in the coexisting phase}
\label{sec:local spin dynamics}
{\renewcommand{\figurename}{}\renewcommand{\thefigure}{Fig. 9 (Color online)} 
\begin{figure}
\begin{center}
\vspace{0.8cm}
  \includegraphics[angle=0,width=0.55\textwidth]{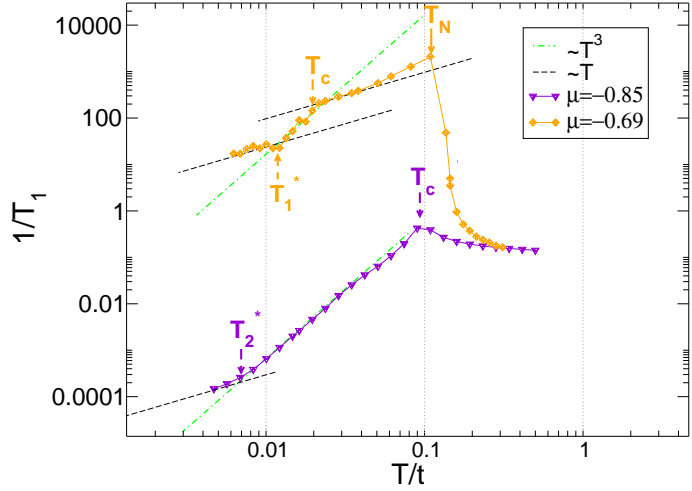}
\end{center}
\vspace{-0.3cm}
\caption{The NMR spin-lattice relaxation rate $1/T_1$ as a function of temperature for the fixed chemical potential $\mu=-0.69$ and $-0.85$ in units of the nearest neighbor hopping coefficient $t$. Figure 1 shows the trajectories in the phase diagram for the two cases, which end up in the coexistence (CO) and the superconducting (SC) phases at $T=0$, respectively.}
\label{T1}
\end{figure}
}

The spin-lattice relaxation rate $1/T_1$ can be written as
\begin{equation}
\frac{1}{T_1}=\gamma^2 A^2\int_{0}^{\infty}\langle \hat{S}_+(t)\hat{S}_-(0)\rangle e^{i\omega_0 t} dt
\label{eq_T1}
\end{equation}
where $\gamma$ is the gyromagnetic ratio of the nuclear spin $\hat{\bf I}$ and $A$ is the hyperfine coupling constant between the nuclear and the quasi-particle spins $\hat{\bf S}$, given by $\mathcal{H}_{hyp}=A\hat{\bf I}\cdot \hat{\bf S}$. 
 The nuclear Larmor frequency $\omega_0$ can be set to be zero in the limit $\omega_0 \tau \ll 1$ where $\tau$ is the fluctuation time scale of the hyperfine field.

 The expression in Eq.~(\ref{eq_T1}) can be rewritten in terms of the dynamical susceptibility 
through the fluctuation-dissipation theorem as
~\cite{Moriya1965}
\begin{equation}
\frac{1}{T_1}=\frac{\gamma^2 k_B T}{\hbar}\lim_{\omega\rightarrow 0}\frac{1}{N_0}\sum_{{\bf q}} A^2({\bf q})\frac{\Im\chi^{\perp}_{RPA}({\bf q},\omega+i0^+)}{\omega}
\label{eq_T1_2}
\end{equation}

 Figure 9 shows the spin-lattice relaxation rate $1/T_1$ calculated 
in systems of SC and CO phases with fixed chemical potentials 
$\mu/t=-0.85$ and $-0.69$, respectively. 
For each case, the trajectory in the $T$-$n_e$ phase diagram are shown in Fig. 1. For $\mu/t=-0.85$, a system enters the SC phase around $T_c/t=0.09$. 
Therefore, the $T^3$-behavior is due to 
the dispersion relation of quasi-particle $\sqrt{\varepsilon_{\bf k}^2+\Delta_{\bf k}^2}$ in the superconducting state with $d_{x^2-y^2}$ symmetry. 
On the other hand, for $\mu/t=-0.69$, it undergoes successive transitions at $T_N/t=0.132$ and $T_c/t=0.0196$ to arrive at the CO phase. 
Similarly, the dispersion relation of a quasi-particle branch $\sqrt{\tilde{E}_{\bf k-}^2+\Delta_{\bf k}^2}$ is responsible for the $T^3$-temperature dependence below $T_c$ for $\mu/t=-0.69$, while the NQR relaxation rate shows the metalic behavior due to the existence of the Fermi surface in the temperature range $T_c<T<T_N$. 
For both cases, $1/T_1$ behaves as $T$-linear in the lowest temperature region due to the finite imaginary part $\delta/t=0.001$ in the process of analytical continuation i$\Omega_n\rightarrow\omega+{\rm i}\delta$. This point is discussed in the following. These behaviors shown in Fig. 9 are entirely consistent with NMR experimental data in the superconducting and coexistent phases of CeRhIn$_5$. 
The crossover temperature to $1/T_1\propto T$ will be determined by the effective coupling constant of impurity scattering. In our calculation, the $T$-linear behavior is originated from a small but finite imaginary part $\delta/t=0.001$ of frequency in the analytic continuation {\color{darkblue} ${\rm i}\Omega_n\rightarrow\omega+{\rm i} \delta$~\cite{comment1}.} On the other hand, considering the impurity scattering of quasi-particle, the self-energy correction will be given by $\Sigma_{imp}=-{\rm i}\delta_{imp}=-{\rm i}\pi n_{imp}|u_{\bf k-k'}^{eff}|^2\delta(E_{{\bf k}-}-E_{{\bf k'}-})$ in the present case, where $n_{imp}$ and $u_{\bf q}^{eff}$ are the impurity density and the Fourier transformed effective impurity potential, respectively. 
If the impurity is included in the sample, the residual density of states around the nodal points of the superconducting gap is induced by the impurity scattering. Then, the Korringa behavior is expected from the residual density of {\color{darkblue} states~\cite{Hotta1993,Hirschfeld1998,Bang2004}.} Here, considering the vertex correction by the spin fluctuation for the impurity scattering, the magnitude of effective impurity potential is enhanced~\cite{Miyake2002}. The crossover temperature measured in the NQR experiment is almost independent of pressure varing from 1.82 to 2.35GPa, where the ground state changes from the coexistent state to the superconducting state by increasing pressure. Thus, the contribution of the vertex correction by spin fluctuations will be small, because the pressure of 2.05GPa is really close to the critical pressure of the antiferromagnetic quantum critical point. 

\section{Conclusion}
\label{sec:conclusion}
We have formulated the spin-excitation spectrum in the coexistent phase of antiferromagnetism and $d$-wave superconductivity within RPA to calculate the low energy spin excitation. It has been shown that additional low energy spin excitations with incommensurate momenta develop around the phase boundary between the antiferromagnetic phase and the coexistent phase. This implies that the corresponding incommensurate antiferromagnetic phases can appear between the antiferromagnetic and the coexistent phases~\cite{Aso2009, Park2012}. {\color{darkblue} We have suggested that the stability of the commensurate
coexistent state is related with the spin resonance mode shown in the
coexistent phase, as the spin resonance stabilizes the $d$-wave
superconducting state in high-$T_c$ cuprates and CeCoIn$_5$.}
The NMR relaxation rate has also been calculated. The calculated result for the system with the coexistent ground state could explain well the temperature dependence of NQR data in the coexistent phase of CeRhIn$_5$. 

\section{Acknowledgement}
 H.-J.Lee thanks KIAS for providing computing resources (KIAS CAC Linux Cluster System). T.T. acknowledges the Max Planck Society and Korea Ministry of Education, Science and Technology for the joint support of the Independent Junior Research Group at the Asia Pacific Center for Theoretical Physics.

\appendix
\section{Mean-Field Equations for the Coexistence of AF and SC}
\label{app:dispersion}
In the CO phase with $\Delta\neq 0$ and $m_s\neq 0$, the mean-field Dyson equations in Eq.~(\ref{dyson_eq0-1})$\sim$(\ref{dyson_eq1}) involve the four Green's functions, 
\begin{eqnarray}
G({\bf k}, \tau)&=&-\frac{1}{2}\sum_\sigma  \langle T_{\tau}\left[ c_{{\bf k}\sigma}(\tau) c^\dagger_{{\bf k}\sigma}(0)\right]\rangle \nonumber\\
F^{z\bf Q}({\bf k}, \tau)&=&-\frac{1}{2}\sum_\sigma  \sigma\langle T_{\tau}\left[ c_{{\bf k+Q}\sigma}(\tau) c^\dagger_{{\bf k}\sigma}(0)\right]\rangle \nonumber\\
F^{s\dagger}({\bf k}, \tau)&=&-\frac{1}{2}\sum_\sigma  \sigma\langle T_{\tau}\left[ c_{{\bf -k}-\sigma}^{\dagger}(\tau) c_{{\bf k}\sigma}^{\dagger}(0)\right]\rangle \nonumber\\
F^{t{\bf Q}\dagger}({\bf k}, \tau)&=&-\frac{1}{2}\sum_\sigma  \langle T_{\tau}\left[ c^\dagger_{-{\bf k}-\sigma}(\tau) c^\dagger_{{\bf k}\sigma}(0)\right].
\label{greenftn1}
\end{eqnarray}
The momentm average of $F^{z\bf Q}({\bf k}, \tau)$ and $F^{s\dagger}({\bf k}, \tau)=F^{s}({\bf k}, \tau)^*$ corresponds to the expectation value of the order parameter for the staggered magnetization and the $d_{x^2-y^2}$-wave SC gap,
\begin{eqnarray}
N_0 m_s&=&\sum_{\bf k^\prime}F^{z\bf Q}({\bf k^\prime}, \tau=0),\nonumber\\
{\Delta}_{\bf k}&=&
\sum_{\bf k^\prime}V_{\bf k,k^\prime}F^s({\bf k^\prime}, \tau=0).
\end{eqnarray}
The last term $F^{t{\bf Q}\dagger}({\bf k}, \tau)$ makes a coupling between $F^{z\bf Q}({\bf k}, \tau)$ and $F^{s}({\bf k}, \tau)$ as shown in Eq.~(\ref{dyson_eq1}).

After the Fourier transformation to the imaginary frequency, the generalized Dyson-Gorkov equation is written as
\begin{eqnarray}
\left[\begin{array}{c}
    G({\bf k}, {\rm i}\omega_n)\\
F^{z\bf Q}({\bf k},{\rm i}\omega_n )\\
F^{s\dagger}({\bf k},{\rm i}\omega_n )\\
F^{t{\bf Q}\dagger}({\bf k},{\rm i}\omega_n )
\end{array}\right]
&=&\hat{A}^{-1}\left[\begin{array}{c}
    G^{(0)}({\bf k},{\rm i}\omega_n)\\
    0 \\
    0\\
    0
  \end{array}\right]
\end{eqnarray}
with 
\begin{eqnarray}
\hat{A}&=&\left[\begin{array}{cccc}
    1 &     G^{(0)}(k)Um_s &    G^{(0)}(k)\Delta_{\bf k}  &  0 \\
    G^{(0)}(k+Q)Um_s &  1& 0& G^{(0)}(k+Q)\Delta_{\bf k+Q}  \\
    -G^{(0)}(-k)\Delta^*_{\bf k} &  0 & 1 &   -G^{(0)}(-k)U m_s\\
    0 &     -G^{(0)}(-k-Q)\Delta^*_{\bf k+Q} &    -G^{(0)}(-k-Q)Um_s &  1. 
\end{array}\right]
\end{eqnarray}
where $G^{(0)}(k)^{-1}={\rm i}\omega_n-\varepsilon_{\bf k}$ and 
$G^{(0)}(k+Q)^{-1}={\rm i}\omega_n-\varepsilon_{\bf k+Q}$.
The Green's function is given by using the cofactor expansion,
\begin{eqnarray}
    G({\bf k}, {\rm i}\omega_n)&=&\frac{A_{11}}{\det\hat{A}} G^{(0)}({\bf k}, {\rm i}\omega_n),\nonumber\\
F^{z\bf Q}({\bf k},{\rm i}\omega_n )&=&\frac{A_{12}}{\det\hat{A}} G^{(0)}({\bf k}, {\rm i}\omega_n),\nonumber\\
F^{s\dagger}({\bf k},{\rm i}\omega_n )&=&\frac{A_{13}}{\det\hat{A}} G^{(0)}({\bf k}, {\rm i}\omega_n),\nonumber\\
F^{t{\bf Q}\dagger}({\bf k},{\rm i}\omega_n )&=&\frac{A_{14}}{\det\hat{A}} G^{(0)}({\bf k}, {\rm i}\omega_n),
\end{eqnarray}
where $\det \hat{A}$ is obtained 
by using the cofactor $A_{1j}$ $(j=1,4)$. 
If the superconducting gap function satisfies $\Delta_{\bf k+Q}=-\Delta_{\bf k}$ like the case of $\Delta_{\bf k}=\Delta(\cos k_x -\cos k_y)$, 
the secular equation 
\begin{equation}
|G^{(0)}(k)G^{(0)}(k+Q)|^{-2}\det{\hat{A}}=\left[({\rm i}\omega_n)^2-\tilde{E}^2_{\bf k-}\right]\left[({\rm i}\omega_n)^2-\tilde{E}^2_{\bf k+}\right]=0
\end{equation}
gives rather simple two eigenvalues $\tilde{E}_{\bf k\pm}$ as
\begin{equation}
\tilde{E}^2_{\bf k\pm}=\left[\frac{\varepsilon_{\bf k}+\varepsilon_{\bf k+Q}}{2}\pm\sqrt{\frac{(\varepsilon_{\bf k}-\varepsilon_{\bf k+Q})^2}{4}+(Um_s)^2}\right]^2+\Delta_{{\bf k}}^2.
\label{dispersion_CO2}
\end{equation}
Then, the Green's functions are given as
\begin{eqnarray}
    D({\bf k},{\rm i}\omega_n)G({\bf k}, {\rm i}\omega_n)&=&(-{\rm i}\omega_n-\varepsilon_{\bf k})\left[-({\rm i}\omega_n)^2+\varepsilon^2_{\bf k+Q} \right]-({\rm i}\omega_n-\varepsilon_{\bf k+Q})(Um_s)^2+({\rm i}\omega_n+\varepsilon_{\bf k})|\Delta_{\bf k}|^2\nonumber\\
D({\bf k},{\rm i}\omega_n)F^{z\bf Q}({\bf k},{\rm i}\omega_n )&=&(-Um_s)(-{\rm i}\omega_n-\varepsilon_{\bf k})(-{\rm i}\omega_n-\varepsilon_{\bf k+Q})+(Um_s)^3-(-Um_s)|\Delta_{\bf k}|^2\nonumber\\
D({\bf k},{\rm i}\omega_n)F^{s}({\bf k},{\rm i}\omega_n )&=&\Delta^*_{\bf k}\left[-({\rm i}\omega_n)^2+\varepsilon^2_{\bf k+Q}+(Um_s)^2+|\Delta_{\bf k}|^2 \right]\nonumber\\
D({\bf k},{\rm i}\omega_n)F^{t{\bf Q}\dagger}({\bf k},{\rm i}\omega_n )&=&(-Um_s\Delta^*_{\bf k})(\varepsilon_{\bf k}+\varepsilon_{\bf k+Q}).
\label{greenftn1}
\end{eqnarray}
with $D({\bf k},{\rm i}\omega_n)=\left[({\rm i}\omega_n)^2-\tilde{E}^2_{\bf k-}\right]\left[({\rm i}\omega_n)^2-\tilde{E}^2_{\bf k+}\right]$. 
Using the Green's functions in Eq.~(\ref{greenftn1}), the following mean-field equations are obtained
\begin{eqnarray}
\frac{n_e}{2}
&=&\frac{1}{N_0}\sum_{\bf k}^\prime \left[1-\frac{E_{{\bf k}+}}{2\tilde{E}_{{\bf k}+}}\tanh\frac{\tilde{E}_{{\bf k}+}}{2T}-\frac{E_{{\bf k}-}}{2\tilde{E}_{{\bf k}-}}\tanh\frac{\tilde{E}_{{\bf k}-}}{2T}\right],\nonumber \\
{m_s}&=&\frac{1}{N_0}\sum_{\bf k}^\prime \frac{Um_s}{\sqrt{t^2_{1\bf k}+(Um_s)^2}}\left[\frac{E_{{\bf k}+}}{2\tilde{E}_{{\bf k}+}}\tanh\frac{\tilde{E}_{{\bf k}+}}{2T}-\frac{E_{{\bf k}-}}{2\tilde{E}_{{\bf k}-}}\tanh\frac{\tilde{E}_{{\bf k}-}}{2T}\right],
\label{app:orderparameter1}\\
{\Delta}_{\bf k}
&=&\sum_{\bf k^\prime}^\prime V_{\bf k,k^\prime}\left[\frac{\Delta_{{\bf k}^\prime}}{2\tilde{E}_{{\bf k^\prime}+}}\tanh\frac{\tilde{E}_{{\bf k^\prime}+}}{2T}+\frac{\Delta_{{\bf k^\prime}}}{2\tilde{E}_{{\bf k^\prime}-}}\tanh\frac{\tilde{E}_{{\bf k^\prime}-}}{2T}\right], \nonumber \label{app:orderparameter3}
\end{eqnarray}
where $\sum_{\bf k}^\prime$ is the momentum summation within the magnetic Brillouin zone.  

\section{Calculation of spin fluctuaion spectrum in the presence of AF and SC long-range order}
\label{app:spin fluctuation}
In the CO phase, 
there are two sites 
in the magnetic unit cell due to the staggered moment $m_s$. Then, the area of magnetic Brillouin zone is a half of the original one, and the ordering wave vector {\bf Q} becomes a reciprocal lattice vector. 
Therefore, 
momenta of initial and final states of relevant spin fluctuation can differ by the reciprocal lattice vector as the Umklapp process. 
Since this fact is independent of components of spin operators, 
we introduce a generalized form of the fluctuation with $\alpha, \beta=0, x,y,z,$ components, as follows,
\begin{equation}
\chi^{\alpha\beta}({\bf q},{\bf q^\prime},{\rm i} \Omega_n)=\int_0^{1/T} d\tau e^{{\rm i} \Omega_n \tau}\langle T_\tau\left[ \delta S^\alpha_{\bf q}(\tau)\delta S^\beta_{-\bf q^\prime}(0)\right]\rangle,
\end{equation}
where $\alpha,\beta=0$ corresponds to the charge operator. 
Furthermore, when the direction of staggered moment is parallel to $z$-direction, the fluctuation matrix is decomposed into two $4\times 4$ matrices formed by $(0,z)$-components and $(x,y)$-components. 


The transversal components accompanying the contribution of spin wave mode 
are given by 
\begin{eqnarray}
 &&\chi^{xx}({\bf q},{\bf q'},{\rm i}\Omega_n)
  =\sum_{\sigma,\sigma'}
  \tilde{\chi}^{\sigma\sigma'}({\bf q},{\bf q'},{\rm i}\Omega_n),\\
 &&\chi^{yy}({\bf q},{\bf q'},{\rm i}\Omega_n)
  =\sum_{\sigma,\sigma'}\sigma\sigma'
  \tilde{\chi}^{\sigma\sigma'}({\bf q},{\bf q'},{\rm i}\Omega_n),\\
 &&\chi^{xy}({\bf q},{\bf q'},{\rm i}\Omega_n)
  ={\rm i}\sum_{\sigma,\sigma'}\sigma'
  \tilde{\chi}^{\sigma\sigma'}({\bf q},{\bf q'},{\rm i}\Omega_n),\\
 &&\chi^{yx}({\bf q},{\bf q'},{\rm i}\Omega_n)
  =-{\rm i}\sum_{\sigma,\sigma'}\sigma
  \tilde{\chi}^{\sigma\sigma'}({\bf q},{\bf q'},{\rm i}\Omega_n),
\end{eqnarray}
with
\begin{eqnarray}
 \tilde{\chi}^{\sigma\sigma'}({\bf q},{\bf q'},{\rm i}\Omega_n)
  =\int^{1/T}_{0}d\tau\hspace{1mm}e^{{\rm i}\Omega_n\tau}
   \frac{1}{4}\sum_{\bf k,k'}
   \langle T_{\tau}[c_{{\bf k}\sigma}^{\dagger}(\tau)c_{{\bf k+q}-\sigma}(\tau)
                    c_{{\bf k'+q'}-\sigma'}^{\dagger}c_{{\bf 'q}\sigma'}]
   \rangle.
\end{eqnarray}
Through the analytical continuation to the real axis, 
the transversal susceptibility matrix obtained within RPA is given by
\begin{eqnarray}
 \hat{\chi}^{\perp}_{RPA}({\bf q},\omega+{\rm i}\delta)
  =[\hat{1}-2\hat{\bar{\chi}}^{\perp}({\bf q},\omega+{\rm i}\delta)
            \hat{U}^{\perp}]^{-1}
   \hat{\bar{\chi}}^{\perp}({\bf q},\omega+{\rm i}\delta),
\end{eqnarray}
with the interaction matrix
\begin{eqnarray}
 \hat{U}^{\perp}=
 \left[
  \begin{array}{cccc}
   U & 0 & 0 & 0\\
   0 & U & 0 & 0\\
   0 & 0 & U & 0\\
   0 & 0 & 0 & U\\
  \end{array}
  \right],
\end{eqnarray}
and the irreducible transversal susceptibility
\begin{eqnarray}
 \hat{\bar{\chi}}^{\perp}({\bf q},\omega+{\rm i}\delta)=
 \left[
  \begin{array}{cccc}
   \bar{\chi}^{xx}({\bf q},{\bf q}) & 0 & 
    0 & \bar{\chi}^{xy}({\bf q},{\bf q+Q})\\
   0 & \bar{\chi}^{xx}({\bf q+Q},{\bf q+Q}) & 
    \bar{\chi}^{xy}({\bf q+Q},{\bf q}) & 0\\
   0 & \bar{\chi}^{yx}({\bf q},{\bf q+Q}) & 
    \bar{\chi}^{yy}({\bf q},{\bf q}) & 0\\
   \bar{\chi}^{yx}({\bf q+Q},{\bf q}) & 0 & 
    0 & \bar{\chi}^{yy}({\bf q+Q},{\bf q+Q})\\
  \end{array}
  \right],
\end{eqnarray}
where the frequency dependence of every matrix element is omitted 
in the right hand side. 
The matrix elements of irreducible transversal susceptibility 
are given with $\tilde{\omega}=\omega+{\rm i}\delta$, as follows,
\begin{eqnarray}
 &&\bar{\chi}^{xx}({\bf q},{\bf q},\tilde{\omega})=
 \bar{\chi}^{yy}({\bf q},{\bf q},\tilde{\omega})
\nonumber\\
 &=&\frac{1}{2N_0}\sum'_{\bf k}
   [\alpha_{\bf k,k+q}^{-}
    \{
     S^{I++}_{\bf k,k+q}
     \frac{f(\tilde{E}_{{\bf k+q}+})-f(\tilde{E}_{{\bf k}+})}
          {\tilde{E}_{{\bf k}+}-\tilde{E}_{{\bf k+q}+}+\tilde{\omega}}
\nonumber\\
 &&+\frac{1}{2}
     S^{II++}_{\bf k,k+q}
     \left(
      \frac{1-f(\tilde{E}_{{\bf k+q}+})-f(\tilde{E}_{{\bf k}+})}
           {\tilde{E}_{{\bf k}+}+\tilde{E}_{{\bf k+q}+}+\tilde{\omega}}
     +\frac{1-f(\tilde{E}_{{\bf k+q}+})-f(\tilde{E}_{{\bf k}+})}
           {\tilde{E}_{{\bf k}+}+\tilde{E}_{{\bf k+q}+}-\tilde{\omega}}
     \right)
\nonumber\\
 &&+S^{I--}_{\bf k,k+q}
     \frac{f(\tilde{E}_{{\bf k+q}-})-f(\tilde{E}_{{\bf k}-})}
          {\tilde{E}_{{\bf k}-}-\tilde{E}_{{\bf k+q}-}+\tilde{\omega}}
\nonumber\\
 &&+\frac{1}{2}
     S^{II--}_{\bf k,k+q}
     \left(
      \frac{1-f(\tilde{E}_{{\bf k+q}-})-f(\tilde{E}_{{\bf k}-})}
           {\tilde{E}_{{\bf k}-}+\tilde{E}_{{\bf k+q}-}+\tilde{\omega}}
     +\frac{1-f(\tilde{E}_{{\bf k+q}-})-f(\tilde{E}_{{\bf k}-})}
           {\tilde{E}_{{\bf k}-}+\tilde{E}_{{\bf k+q}-}-\tilde{\omega}}
     \right)
    \}
\nonumber\\
 &&+\beta_{\bf k,k+q}^{-}
    \{
     S^{I+-}_{\bf k,k+q}
     \left(
      \frac{f(\tilde{E}_{{\bf k+q}-})-f(\tilde{E}_{{\bf k}+})}
           {\tilde{E}_{{\bf k}+}-\tilde{E}_{{\bf k+q}-}+\tilde{\omega}}
     +\frac{f(\tilde{E}_{{\bf k+q}-})-f(\tilde{E}_{{\bf k}+})}
           {\tilde{E}_{{\bf k}+}-\tilde{E}_{{\bf k+q}-}-\tilde{\omega}}
     \right)
\nonumber\\
  &&
    +S^{II+-}_{\bf k,k+q}
     \left(
      \frac{1-f(\tilde{E}_{{\bf k+q}-})-f(\tilde{E}_{{\bf k}+})}
           {\tilde{E}_{{\bf k}+}+\tilde{E}_{{\bf k+q}-}+\tilde{\omega}}
     +\frac{1-f(\tilde{E}_{{\bf k+q}-})-f(\tilde{E}_{{\bf k}+})}
           {\tilde{E}_{{\bf k}+}+\tilde{E}_{{\bf k+q}-}-\tilde{\omega}}
     \right)
     \}],
   \label{chi_perp}
\end{eqnarray}
\begin{eqnarray}
 &&\bar{\chi}^{xy}({\bf q},{\bf q\pm Q},\tilde{\omega})=
 -\bar{\chi}^{yx}({\bf q},{\bf q\pm Q},\tilde{\omega})
\nonumber\\
 &=&\frac{\rm -i}{2N_0}\sum'_{\bf k}
   [+\gamma_{\bf k,k+q}^{-}
    \{
     T^{I++}_{\bf k,k+q}
     \frac{f(\tilde{E}_{{\bf k+q}+})-f(\tilde{E}_{{\bf k}+})}
          {\tilde{E}_{{\bf k}+}-\tilde{E}_{{\bf k+q}+}+\tilde{\omega}}
\nonumber\\
 &&+\frac{1}{2}
     T^{II++}_{\bf k,k+q}
     \left(
      \frac{1-f(\tilde{E}_{{\bf k+q}+})-f(\tilde{E}_{{\bf k}+})}
           {\tilde{E}_{{\bf k}+}+\tilde{E}_{{\bf k+q}+}+\tilde{\omega}}
     -\frac{1-f(\tilde{E}_{{\bf k+q}+})-f(\tilde{E}_{{\bf k}+})}
           {\tilde{E}_{{\bf k}+}+\tilde{E}_{{\bf k+q}+}-\tilde{\omega}}
     \right)
\nonumber\\
 &&
    -T^{I--}_{\bf k,k+q}
     \frac{f(\tilde{E}_{{\bf k+q}-})-f(\tilde{E}_{{\bf k}-})}
          {\tilde{E}_{{\bf k}-}-\tilde{E}_{{\bf k+q}-}+\tilde{\omega}}
\nonumber\\
 &&-\frac{1}{2}
     T^{II--}_{\bf k,k+q}
     \left(
      \frac{1-f(\tilde{E}_{{\bf k+q}-})-f(\tilde{E}_{{\bf k}-})}
           {\tilde{E}_{{\bf k}-}+\tilde{E}_{{\bf k+q}-}+\tilde{\omega}}
     -\frac{1-f(\tilde{E}_{{\bf k+q}-})-f(\tilde{E}_{{\bf k}-})}
           {\tilde{E}_{{\bf k}-}+\tilde{E}_{{\bf k+q}-}-\tilde{\omega}}
     \right)
    \}
\nonumber\\
 &&+\gamma_{\bf k,k+q}^{+}
    \{
     T^{I+-}_{\bf k,k+q}
     \left(
      \frac{f(\tilde{E}_{{\bf k+q}-})-f(\tilde{E}_{{\bf k}+})}
           {\tilde{E}_{{\bf k}+}-\tilde{E}_{{\bf k+q}-}+\tilde{\omega}}
     -\frac{f(\tilde{E}_{{\bf k+q}-})-f(\tilde{E}_{{\bf k}+})}
           {\tilde{E}_{{\bf k}+}-\tilde{E}_{{\bf k+q}-}-\tilde{\omega}}
     \right)
\nonumber\\
  &&
    +T^{II+-}_{\bf k,k+q}
     \left(
      \frac{1-f(\tilde{E}_{{\bf k+q}-})-f(\tilde{E}_{{\bf k}+})}
           {\tilde{E}_{{\bf k}+}+\tilde{E}_{{\bf k+q}-}+\tilde{\omega}}
     -\frac{1-f(\tilde{E}_{{\bf k+q}-})-f(\tilde{E}_{{\bf k}+})}
           {\tilde{E}_{{\bf k}+}+\tilde{E}_{{\bf k+q}-}-\tilde{\omega}}
     \right)
     \}],
\end{eqnarray}
with coherence factors due to antiferromagnetism
\begin{eqnarray}
&&\alpha^{\pm}_{\bf k,k+q}=\frac{1}{2}\left( 1+ \frac{t_{1\bf k}t_{1\bf k+q}\pm (Um)^2}{\sqrt{t_{1\bf k}^2+(Um)^2}\sqrt{t_{1\bf k+q}^2+(Um)^2}}\right),\nonumber\\
&&\beta^{\pm}_{\bf k,k+q}=\frac{1}{2}\left( 1- \frac{t_{1\bf k}t_{1\bf k+q}\pm (Um)^2}{\sqrt{t_{1\bf k}^2+(Um)^2}\sqrt{t_{1\bf k+q}^2+(Um)^2}}\right),\nonumber\\
&&\gamma^{\pm}_{\bf k,k+q}=\frac{1}{2}\left( \frac{Um}{\sqrt{t_{1\bf k}^2+(Um)^2}}\pm\frac{Um}{\sqrt{t_{1\bf k+q}^2+(Um)^2}}\right),\label{cohef3}
\end{eqnarray}
and coherence factors due to superconductivity
\begin{eqnarray}
&&S^{I\xi\eta}_{\bf k,k+q}
 =\frac{1}{2}\left(1+\frac{E_{{\bf k}\xi}E_{{\bf k+q}\eta}
                          +\Delta_{\bf k}\Delta_{\bf k+q}}
                          {\tilde{E}_{{\bf k}\xi}\tilde{E}_{{\bf k+q}\eta}}
             \right),\\
&&S^{II\xi\eta}_{\bf k,k+q}
 =\frac{1}{2}\left(1-\frac{E_{{\bf k}\xi}E_{{\bf k+q}\eta}
                          +\Delta_{\bf k}\Delta_{\bf k+q}}
                          {\tilde{E}_{{\bf k}\xi}\tilde{E}_{{\bf k+q}\eta}}
             \right),\\
&&T^{I\xi\eta}_{\bf k,k+q}
 =\frac{1}{2}\left(\frac{E_{{\bf k}\xi}}{\tilde{E}_{{\bf k}\xi}}
                  +\frac{E_{{\bf k+q}\eta}}{\tilde{E}_{{\bf k+q}\eta}}
             \right),\hspace{10mm}
  T^{II\xi\eta}_{\bf k,k+q}
 =\frac{1}{2}\left(\frac{E_{{\bf k}\xi}}{\tilde{E}_{{\bf k}\xi}}
                  -\frac{E_{{\bf k+q}\eta}}{\tilde{E}_{{\bf k+q}\eta}}
             \right).
\end{eqnarray}
These matrix elements of transversal irreducible susceptibilities 
are calculated through the irreducible susceptibility of 
$\tilde{\chi}^{\sigma\sigma'}({\bf q},{\bf q'},{\rm i}\Omega_n)$, 
as follows,
\begin{eqnarray}
 \bar{\tilde{\chi}}^{\sigma\sigma'}({\bf q},{\bf q'},{\rm i}\Omega_n)
  &=&-\delta_{\sigma,\sigma'}\frac{T}{4N_0}\sum_{\bf k}\sum_l
  [\delta_{\bf q-q'}
   \{G({\bf k},{\rm i}\omega_l)
     G({\bf k+q},{\rm i}\omega_l+{\rm i}\Omega_n)\nonumber\\
  &&\hspace{35mm}-F^{z{\bf Q}}({\bf k},{\rm i}\omega_l)
   F^{z{\bf -Q}}({\bf k+q+Q},{\rm i}\omega_l+{\rm i}\Omega_n)\}\nonumber\\
  &&\hspace{24mm}-\delta_{\bf q-q'+Q}\sigma
   \{G({\bf k},{\rm i}\omega_l)
     F^{z{\bf -Q}}({\bf k+q+Q},{\rm i}\omega_l+{\rm i}\Omega_n)\nonumber\\
  &&\hspace{43mm}-F^{z{\bf Q}}({\bf k},{\rm i}\omega_l)
     G({\bf k+q},{\rm i}\omega_l+{\rm i}\Omega_n)\}\nonumber\\
  &&\hspace{24mm}+\delta_{\bf q-q'}
   \{F^s({\bf k},{\rm i}\omega_l)
     F^{s\dagger}({\bf k+q},{\rm i}\omega_l+{\rm i}\Omega_n)\nonumber\\
  &&\hspace{35mm}-F^{t{\bf Q}}({\bf k},{\rm i}\omega_l)
  F^{t{\bf Q}\dagger}({\bf k+q},{\rm i}\omega_l+{\rm i}\Omega_n)\}\nonumber\\
  &&\hspace{24mm}+\delta_{\bf q-q'+Q}\sigma\{
     F^{t{\bf Q}}({\bf k},{\rm i}\omega_l)
     F^{s\dagger}({\bf k+q+Q},{\rm i}\omega_l+{\rm i}\Omega_n)\nonumber\\
  &&\hspace{43mm}-
     F^s({\bf k},{\rm i}\omega_l)
     F^{t{\bf Q}\dagger}({\bf k+q-Q},{\rm i}\omega_l+{\rm i}\Omega_n)\}].
\end{eqnarray}
Similarly, 
the dynamical susceptibility of charge and $z$-component of spin 
are estimated within RPA as
\begin{eqnarray}
 &&\hat{\chi}^{\parallel}_{RPA}({\bf q},\omega+{\rm i}\delta)
  =[\hat{1}-2\hat{\bar{\chi}}^{\parallel}({\bf q},\omega+{\rm i}\delta)
            \hat{U}^{\parallel}]^{-1}
   \hat{\bar{\chi}}^{\parallel}({\bf q},\omega+{\rm i}\delta),
\end{eqnarray}
with the interaction matrix
\begin{eqnarray}
 \hat{U}^{\parallel}=
 \left[
  \begin{array}{cccc}
   -U & 0 & 0 & 0\\
   0 & -U & 0 & 0\\
   0 & 0 & U & 0\\
   0 & 0 & 0 & U\\
  \end{array}
 \right],
\end{eqnarray}
and the irreducible spin-susceptibility
\begin{eqnarray}
 \hat{\bar{\chi}}^{\parallel}({\bf q},\omega+{\rm i}\delta)=
 \left[
  \begin{array}{cccc}
   \bar{\chi}^{00}({\bf q},{\bf q}) & 0 & 
    0 & \bar{\chi}^{0z}({\bf q},{\bf q+Q})\\
   0 & \bar{\chi}^{00}({\bf q+Q},{\bf q+Q}) & 
    \bar{\chi}^{0z}({\bf q+Q},{\bf q}) & 0\\
   0 & \bar{\chi}^{z0}({\bf q},{\bf q+Q}) & 
    \bar{\chi}^{zz}({\bf q},{\bf q}) & 0\\
   \bar{\chi}^{z0}({\bf q+Q},{\bf q}) & 0 & 
    0 & \bar{\chi}^{zz}({\bf q+Q},{\bf q+Q})\\
  \end{array}
 \right].
\end{eqnarray}
These matrix elements are given as
\begin{eqnarray}
 &&\left[
   \begin{array}{c}
    \bar{\chi}^{00}({\bf q},{\bf q},\tilde{\omega})\\
    \bar{\chi}^{zz}({\bf q},{\bf q},\tilde{\omega})
   \end{array}
   \right]
\nonumber\\
 &=&\frac{1}{2N_0}\sum'_{\bf k}
   [\alpha_{\bf k,k+q}^{+}
    \{
     \left[
       \begin{array}{c}
        C^{I++}_{\bf k,k+q}\\
        S^{I++}_{\bf k,k+q}
       \end{array}
     \right]
     \frac{f(\tilde{E}_{{\bf k+q}+})-f(\tilde{E}_{{\bf k}+})}
          {\tilde{E}_{{\bf k}+}-\tilde{E}_{{\bf k+q}+}+\tilde{\omega}}
\nonumber\\
 &&
   +\frac{1}{2}
     \left[
       \begin{array}{c}
        C^{II++}_{\bf k,k+q}\\
        S^{II++}_{\bf k,k+q}
       \end{array}
     \right]
     \left(
      \frac{1-f(\tilde{E}_{{\bf k+q}+})-f(\tilde{E}_{{\bf k}+})}
           {\tilde{E}_{{\bf k}+}+\tilde{E}_{{\bf k+q}+}+\tilde{\omega}}
     +\frac{1-f(\tilde{E}_{{\bf k+q}+})-f(\tilde{E}_{{\bf k}+})}
           {\tilde{E}_{{\bf k}+}+\tilde{E}_{{\bf k+q}+}-\tilde{\omega}}
     \right)
\nonumber\\
 &&
    +\left[
       \begin{array}{c}
        C^{I--}_{\bf k,k+q}\\
        S^{I--}_{\bf k,k+q}
       \end{array}
     \right]
     \frac{f(\tilde{E}_{{\bf k+q}-})-f(\tilde{E}_{{\bf k}-})}
          {\tilde{E}_{{\bf k}-}-\tilde{E}_{{\bf k+q}-}+\tilde{\omega}}
\nonumber\\
 &&
   +\frac{1}{2}
     \left[
       \begin{array}{c}
        C^{II--}_{\bf k,k+q}\\
        S^{II--}_{\bf k,k+q}
       \end{array}
     \right]
     \left(
      \frac{1-f(\tilde{E}_{{\bf k+q}-})-f(\tilde{E}_{{\bf k}-})}
           {\tilde{E}_{{\bf k}-}+\tilde{E}_{{\bf k+q}-}+\tilde{\omega}}
     +\frac{1-f(\tilde{E}_{{\bf k+q}-})-f(\tilde{E}_{{\bf k}-})}
           {\tilde{E}_{{\bf k}-}+\tilde{E}_{{\bf k+q}-}-\tilde{\omega}}
     \right)
    \}
\nonumber\\
 &&+\beta_{\bf k,k+q}^{+}
    \{
     \left[
       \begin{array}{c}
        C^{I+-}_{\bf k,k+q}\\
        S^{I+-}_{\bf k,k+q}
       \end{array}
     \right]
     \left(
      \frac{f(\tilde{E}_{{\bf k+q}-})-f(\tilde{E}_{{\bf k}+})}
           {\tilde{E}_{{\bf k}+}-\tilde{E}_{{\bf k+q}-}+\tilde{\omega}}
     +\frac{f(\tilde{E}_{{\bf k+q}-})-f(\tilde{E}_{{\bf k}+})}
           {\tilde{E}_{{\bf k}+}-\tilde{E}_{{\bf k+q}-}-\tilde{\omega}}
     \right)
\nonumber\\
  &&+
     \left[
       \begin{array}{c}
        C^{II+-}_{\bf k,k+q}\\
        S^{II+-}_{\bf k,k+q}
       \end{array}
     \right]
     \left(
      \frac{1-f(\tilde{E}_{{\bf k+q}-})-f(\tilde{E}_{{\bf k}+})}
           {\tilde{E}_{{\bf k}+}+\tilde{E}_{{\bf k+q}-}+\tilde{\omega}}
     +\frac{1-f(\tilde{E}_{{\bf k+q}-})-f(\tilde{E}_{{\bf k}+})}
           {\tilde{E}_{{\bf k}+}+\tilde{E}_{{\bf k+q}-}-\tilde{\omega}}
     \right)
     \}],
\label{chi_long}
\end{eqnarray}
\begin{eqnarray}
 &&\left[
   \begin{array}{c}
    \bar{\chi}^{0z}({\bf q},{\bf q\pm Q},\tilde{\omega})\\
    \bar{\chi}^{z0}({\bf q},{\bf q\pm Q},\tilde{\omega})
   \end{array}
   \right]
\nonumber\\
 &=&\frac{1}{2N_0}\sum'_{\bf k}
   [-\gamma_{\bf k,k+q}^{+}
    \{
     \left[
       \begin{array}{c}
        C^{I++}_{\bf k,k+q}\\
        S^{I++}_{\bf k,k+q}
       \end{array}
     \right]
     \frac{f(\tilde{E}_{{\bf k+q}+})-f(\tilde{E}_{{\bf k}+})}
          {\tilde{E}_{{\bf k}+}-\tilde{E}_{{\bf k+q}+}+\tilde{\omega}}
\nonumber\\
 &&+\frac{1}{2}
     \left[
       \begin{array}{c}
        C^{II++}_{\bf k,k+q}\\
        S^{II++}_{\bf k,k+q}
       \end{array}
     \right]
     \left(
      \frac{1-f(\tilde{E}_{{\bf k+q}+})-f(\tilde{E}_{{\bf k}+})}
           {\tilde{E}_{{\bf k}+}+\tilde{E}_{{\bf k+q}+}+\tilde{\omega}}
     +\frac{1-f(\tilde{E}_{{\bf k+q}+})-f(\tilde{E}_{{\bf k}+})}
           {\tilde{E}_{{\bf k}+}+\tilde{E}_{{\bf k+q}+}-\tilde{\omega}}
     \right)
\nonumber\\
 &&-
     \left[
       \begin{array}{c}
        C^{I--}_{\bf k,k+q}\\
        S^{I--}_{\bf k,k+q}
       \end{array}
     \right]
     \frac{f(\tilde{E}_{{\bf k+q}-})-f(\tilde{E}_{{\bf k}-})}
          {\tilde{E}_{{\bf k}-}-\tilde{E}_{{\bf k+q}-}+\tilde{\omega}}
\nonumber\\
 &&-
    \frac{1}{2}
     \left[
       \begin{array}{c}
        C^{II--}_{\bf k,k+q}\\
        S^{II--}_{\bf k,k+q}
       \end{array}
     \right]
     \left(
      \frac{1-f(\tilde{E}_{{\bf k+q}-})-f(\tilde{E}_{{\bf k}-})}
           {\tilde{E}_{{\bf k}-}+\tilde{E}_{{\bf k+q}-}+\tilde{\omega}}
     +\frac{1-f(\tilde{E}_{{\bf k+q}-})-f(\tilde{E}_{{\bf k}-})}
           {\tilde{E}_{{\bf k}-}+\tilde{E}_{{\bf k+q}-}-\tilde{\omega}}
     \right)
    \}
\nonumber\\
 &&-\gamma_{\bf k,k+q}^{-}
    \{
     \left[
       \begin{array}{c}
        C^{I+-}_{\bf k,k+q}\\
        S^{I+-}_{\bf k,k+q}
       \end{array}
     \right]
     \left(
      \frac{f(\tilde{E}_{{\bf k+q}-})-f(\tilde{E}_{{\bf k}+})}
           {\tilde{E}_{{\bf k}+}-\tilde{E}_{{\bf k+q}-}+\tilde{\omega}}
     +\frac{f(\tilde{E}_{{\bf k+q}-})-f(\tilde{E}_{{\bf k}+})}
           {\tilde{E}_{{\bf k}+}-\tilde{E}_{{\bf k+q}-}-\tilde{\omega}}
     \right)
\nonumber\\
  &&+
     \left[
       \begin{array}{c}
        C^{II+-}_{\bf k,k+q}\\
        S^{II+-}_{\bf k,k+q}
       \end{array}
     \right]
     \left(
      \frac{1-f(\tilde{E}_{{\bf k+q}-})-f(\tilde{E}_{{\bf k}+})}
           {\tilde{E}_{{\bf k}+}+\tilde{E}_{{\bf k+q}-}+\tilde{\omega}}
     +\frac{1-f(\tilde{E}_{{\bf k+q}-})-f(\tilde{E}_{{\bf k}+})}
           {\tilde{E}_{{\bf k}+}+\tilde{E}_{{\bf k+q}-}-\tilde{\omega}}
     \right)
     \}],
\end{eqnarray}
and coherence factors $C^{I\xi\eta}_{\bf k,k+q}$ and 
$C^{II\xi\eta}_{\bf k,k+q}$
\begin{eqnarray}
&&C^{I\xi\eta}_{\bf k,k+q}
 =\frac{1}{2}\left(1+\frac{E_{{\bf k}\xi}E_{{\bf k+q}\eta}
                          -\Delta_{\bf k}\Delta_{\bf k+q}}
                          {\tilde{E}_{{\bf k}\xi}\tilde{E}_{{\bf k+q}\eta}}
             \right),\\
&&C^{II\xi\eta}_{\bf k,k+q}
 =\frac{1}{2}\left(1-\frac{E_{{\bf k}\xi}E_{{\bf k+q}\eta}
                          -\Delta_{\bf k}\Delta_{\bf k+q}}
                          {\tilde{E}_{{\bf k}\xi}\tilde{E}_{{\bf k+q}\eta}}
             \right).
\end{eqnarray}
In the absence of the magnetic long-range order ($m=0$), the coherent factor $\gamma^{\pm}_{\bf k,k+q}$ in Eq.~(\ref{cohef3}) vanishes, which makes the off-diagonal components of the susceptibility $\bar{\chi}^{\alpha\beta}({\bf q,q\pm Q},{\rm i}\Omega_n)$ with $\alpha\neq\beta$ zero and only the diagonal component $\bar{\chi}^{\alpha\alpha}({\bf q,q},{\rm i}\Omega_n)$ with $\alpha=x,y,z,0$ remain finite.  In addition, another coherent factors $\beta^{\pm}_{\bf k,k+q}$ also vanishes to drop the last four terms from the diagonal components $\bar{\chi}^{\alpha\alpha}({\bf q,q},{\rm i}\Omega_n)$ with $\alpha=x,y,z,0$ in Eq.~(\ref{chi_perp}) and Eq.~(\ref{chi_long}). Further it makes the other coherent factor $\alpha^{+}_{\bf k,k+q}=\alpha^{-}_{\bf k,k+q}=1$, which recovers the rotational symmetry in spin space and our results of Eq.~(\ref{chi_perp}) and Eq.~(\ref{chi_long}) reduces to the symmetric form of the pure superconducting case~\cite{Yoshikawa1999}.



\begin{thebibliography}{99}
\bibitem{Geibel1991} {C. Geibel, C. Schank, S. Thies, H. Kitazawa, C.D. Bredl, A. B\"{o}hm, M. Rau, A. Granel, R. Caspary, R. Helfrich, U. Ahlheim, G. Weber, and F. Steglich:  Z. Phys. B Condens. Matter {\bf 84} (1991) 1}.
\bibitem{Mathur1998} {N. D. Mathur, F. M. Grosche, S. R. Julian, I. R. Walker, D. M. Freye, R. K. W. Haselwimmer, and G. G. Lonzarich: Nature (London) {\bf 394} (1998) 39}.
\bibitem{Grosche1996} {F. M. Grosche, S. R. Julian, N. D. Mathur, and G. G.. Lonzarich: Physica B {\bf 223$\sim$224} (1996) 50}.
\bibitem{Movshovich1996} {R. Movshovich, T. Graf, D. Mandrus, J. D. Thompson, J. L. Smith, and Z. Fisk: Phys. Rev. B {\bf 53} (1996) 8241}.
\bibitem{Araki2002} {S. Araki, M. Nakashima, R. Settai, T. C. Kobayashi, and Y. ${\rm \bar{O}}$nuki :: J. Phys.: Condens. Matter {\bf 14} (2002) L337}.
\bibitem{Steglich1979} {F. Steglich, J. Aarts, C.D. Bredl, W. Lieke, D. Meschede, W. Franz, and H. Schafer: Phys. Rev. Lett. {\bf 43} (1979) 1892}.
\bibitem{Kawasaki2004} {Y. Kawasaki, K. Ishida, S. Kawasaki, T. Mito, G.-q. Zheng, Y. Kitaoka, C. Geibel and F. Steglich: J. Phys. Soc. Jpn. {\bf 73} (2004) 194}.
\bibitem{Hegger2000} {H. Hegger, C. Petrovic, E. G. Moshopoulou, M. F. Hundley, J. L. Sarrao, Z. Fisk, and J. D. Thompson: Phys. Rev. Lett. {\bf 84} (2000) 4986}.
\bibitem{Muramatsu2001} {T. Muramatsu, N. Tateiwa, T. C. Kobayashi, K. Shimizu, K. Amaya, D. Aoki, H. Shishido, Y. Haga, and Y. ${\rm \bar{O}}$nuki : J. Phys. Soc. Jpn. {\bf 70} (2001) 3362}.
\bibitem{Hashimoto2012} {H. Hashimoto et al., Science {\bf 336} (2012) 1554}.
\bibitem{Bao2000} {W. Bao, P. G. Pagliuso, J. L. Sarrao, J. D. Thompson, Z. Fisk, J. W. Lynn, and R. W. Erwin: Phys. Rev. B {\bf 62} (2000) R14621}.
\bibitem{Llobet2004} {A. Llobet {\it et al.}: Phys. Rev. B {\bf 69} (2004) 024403}.
\bibitem{Majumdar2002} {S. Majumdar, G. Balakrishnan, M. R. Lees, D. McK. Paul, and G. J. Mcintyre: Phys. Rev. B {\bf 66} (2002) 212502}.
\bibitem{Raymond2008} {S. Raymond, G. Knebel, D. Aoki, and J. Flouquet: Phys. Rev. B {\bf 77} (2008) 172502}.
\bibitem{Stock2008} {C. Stock, C. Broholm, J. Hudis, H. J. Kang, and C. Petrovic: Phys. Rev. Lett. {\bf 100} (2008) 087001}.
\bibitem{Yashima2007} {M. Yashima, S. Kawasaki, M. Mukada, Y. Kitaoka, H. Shishido, R. Settai, and  Y. ${\rm \bar{O}}$nuki: Phys. Rev. B {\bf 76} (2007) 020509(R)}.
\bibitem{Aso2009} {N. Aso, K. Ishii, H. Yoshizawa, T. Fujiwara, Y. Uwatoko, G.-F. Chen, N. K. Sato, and K. Miyake: J. of Phys. Soc. Jpn. {\bf 78} (2009) 073703}.
\bibitem{Yashima2009} {M. Yashima, H. Mukada, Y. Kitaoka, H. Shishido, R. Settai, Y. ${\rm \bar{O}}$nuki: Phys. Rev. B {\bf 79} (2009) 214528}.
\bibitem{Petrovic2002} {C. Petrovic, P. G. Pagliuso, M. F. Hundley, R. Movshovich, J. L. Sarrao, J. D. Thompson, Z. Fisk, and P. Monthoux: J. Phys.: Condens. Matter {\bf 13} (2001) L337}.
\bibitem{Murakami1998} {M. Murakami and H. Fukuyama: J. Phys. Soc. Jpn. {\bf 67} (1998) 2784}.
\bibitem{Aperis2008} {A. Aperis, G. Varelogiannis, P. B. Littlewood and B. D. Simons:  J. Phys.: Condens. Matter {\bf 20} (2008) 434235}.
\bibitem{Anderson2009} {B. M. Anderson and P. J. Hirschfeld:  Phys. Rev. B {\bf 79} (2009) 144515}.
\bibitem{Ueda1978} {K. Ueda: J. of Phys. Soc. Jpn. {\bf 44} (1978) 1533}.
\bibitem{Rowe2012} {W. Rowe, J. Knolle, I. Eremin, and P. J. Hirschfeld, arXiv:1207.3834}
\bibitem{Shishido2005} {H. Shishido, R. Settai, H. Harima and Y. ${\rm \bar{O}}$nuki: J. of Phys. Soc. Jpn. {\bf 74} (2005) 1103}.
\bibitem{Sigrist1991} {M. Sigrist and K. Ueda: Rev. Mod. Phys. {\bf 63} (1991) 239}.
\bibitem{Moriya1965} {T.Moriya: J. Phys. Soc. Jpn. {\bf 18} (1965) 516}.
\bibitem{comment1} {The parameter $\delta/t$ is chosen as it is smaller than $T_c/t=0.0196$ and accurate data is obtained in a moderate machine time.}  
\bibitem{Hotta1993} {T. Hotta: J. Phys. Soc. Jpn. {\bf 62} (1993) 274}.
\bibitem{Hirschfeld1998} {P. J. Hirschfeld and P. W\"olfle : Phys. Rev. B {\bf 37} (1998) 83}.
\bibitem{Bang2004} {Y. Bang, M. J. Graf, A. V. Balatsky, and J. D. Thompson : Phys. Rev. B {\bf 69} (2004) 014505}.
\bibitem{Miyake2002} {K. Miyake and O. Narikiyo: J. of Phys. Soc. Jpn. {\bf 71} (2002) 867}.
\bibitem{Park2012} {Tuson Park, H. Lee, I. Martin, X. Lu, V. A. Sidorov, K. Gofryk, F. Ronning, E. D. Bauer and J. D. Thompson: Phys. Rev. Lett. {\bf 108} (2012) 077003}.
\bibitem{Yoshikawa1999} {H. Yoshikawa and T. Moriya: J. of Phys. Soc. Jpn. {\bf 68} (1999) 1340}.












\end{thebibliography}
\end{document}